\documentclass{iopart}
\usepackage[utf8]{inputenc} 
\usepackage{iopams}
\expandafter\let\csname equation*\endcsname\relax
\expandafter\let\csname endequation*\endcsname\relax

\usepackage{mathtools}
\usepackage{amsmath}
\usepackage{amssymb}
\usepackage{amsthm}
\usepackage{cases}

\usepackage{graphicx}
\graphicspath{{./figs/}}
\usepackage[font={small}]{caption}
\usepackage{mwe}
\usepackage{dsfont}
\usepackage{bbm}
\usepackage{cite}

\newcommand{\ue}{\mathrm{e}}

\newcommand{\pa}{\partial}
\renewcommand\Re{\operatorname{Re}}
\renewcommand\Im{\operatorname{Im}}
\newcommand{\la}{\langle}
\newcommand{\ra}{\rangle}

\newcommand{\overbar}[1]{\mkern 1.5mu\overline{\mkern-1.5mu#1\mkern-1.5mu}\mkern 1.5mu}

\newcommand{\R}{\mathbb{R}}

\begin{document}

\title[Bloch oscillations in a Bose-Hubbard chain with single-particle losses]{Bloch oscillations in a Bose-Hubbard chain with single-particle losses}
\author{Bradley Longstaff and Eva-Maria Graefe}
\address{Department of Mathematics, Imperial College London, London SW7 2AZ, United Kingdom}
\begin{abstract}
We theoretically investigate Bloch oscillations in a one-dimensional Bose-Hubbard chain, with single-particle losses from the odd lattice sites described by a Lindblad equation. For a single particle the time evolution of the state is completely determined by a non-Hermitian effective Hamiltonian. We analyse the spectral properties of this Hamiltonian for an infinite lattice and link features of the spectrum to observable dynamical effects, such as frequency doubling in breathing modes. We further consider the case of many particles in the mean-field limit leading to complex nonlinear Schr\"odinger dynamics. Analytic expressions are derived for the generalised nonlinear stationary states and the nonlinear Bloch bands. The interplay of nonlinearity and particle losses leads to peculiar features in the nonlinear Bloch bands, such as the vanishing of solutions and the formation of additional exceptional points. The stability of the stationary states is determined via the Bogoliubov-de Gennes equation and is shown to strongly influence the mean-field dynamics. Remarkably, even far from the mean-field limit, the stability of the nonlinear Bloch bands appears to affect the quantum dynamics. This is demonstrated numerically for a two-particle system.\end{abstract}

\section{Introduction}
In recent years there has been great interest in the dynamics of ultracold atoms and Bose-Einstein condensates (BECs) trapped in optical lattices with a static tilt \cite{Mors06,Wimb05,Zene08,Kolo09}. The tilt can cause the atoms to undergo Bloch oscillations, as has been demonstrated, for example, in remarkable experiments with one and two bosonic atoms \cite{Prei15}. A BEC of weakly-interacting particles also exhibits Bloch oscillations, although this behaviour can dramatically change when the particle interaction strength increases \cite{Witt05}. In a full many-particle treatment, a system of ultracold atoms in a tilted one-dimensional optical lattice with $M=2L+1$ lattice sites can be described by the Bose-Hubbard Hamiltonian 
\begin{equation}
\label{eqn:BH Hamiltonian}
\hat H = -J\sum_{j=-L}^{L-1}\left(\hat{a}_{j+1}^\dag \hat{a}_j + \hat{a}_j^\dag \hat{a}_{j+1}\right) + \frac{U}{2}\sum_{j=-L}^{L} \hat{a}_j^\dag \hat{a}_j^\dag \hat{a}_j \hat{a}_j + F \sum_{j=-L}^{L} j \hat{a}_j^\dag \hat{a}_j.
\end{equation}
The $\hat{a}_j^\dag$ ($\hat{a}_j$) create (annihilate) particles in the ground state of the $j$\textsuperscript{th} site and satisfy the canonical commutation relations
\begin{equation}
[\hat{a}_j,\hat{a}_k^\dag] = \delta_{jk}, \quad [\hat{a}_j,\hat{a}_k] = 0;
\end{equation}
$J$ is a positive coupling constant; $U$ describes pairwise on-site interactions; and $F$ describes the lattice tilt. Throughout this paper we use appropriate dimensionless units with $J=1$ and $\hbar = 1$.

Open quantum systems have become a focus of intense research, motivated in part by the drive towards engineering quantum effects into new technologies. While dissipation and decoherence were once viewed solely as unwanted features, engineered dissipative systems have opened up new opportunities for the control of quantum dynamics. For example, the ability to remove single atoms from specific sites of an optical lattice \cite{Geri08,Wurt09} may be used to control the dynamics of many-body quantum systems. Theoretically, the single-particle losses in such a setup can be described by the Lindblad equation \cite{Angl97,Ruos98,Sina98,Breu02,Dale14}, an approximate master equation that is often valid in atomic, molecular and optical systems. 

The time evolution given by the Lindblad equation can be related to non-Hermitian Schr\"odinger dynamics in various ways. For example, the quantum jump approach consists of an effective non-Hermitian evolution interrupted by irreversible transformations \cite{Dali92,Dum92}. Effective non-Hermitian Hamiltonians have long been used to model open quantum systems and constitute an active research area in their own right. In particular, the class of PT-symmetric Hamiltonians has attracted much attention over the last few decades \cite{Bend18_book}. The differences between the various definitions of PT-symmetry in the literature are subtle. The most common definition in the context of non-Hermitian Hamiltonians refers to the invariance of a Hamiltonian with respect to the combined action of a time-reversal operator $\hat T$, an antilinear operator that squares to plus or minus the identity, and a parity operator $\hat P$, a linear operator that is invariant under $\hat T$ and squares to the identity. The eigenvalues of PT-symmetric Hamiltonians are either real or come in complex conjugate pairs. Loosely speaking, PT-symmetry can be interpreted as a balance of gain and loss in a system. There has been a growing interest in non-Hermitian PT-symmetric quantum dynamics, with many experiments performed in analogous wave systems, such as microwave resonators and optical waveguides \cite{Chris18_book}. Rather than balancing gain and loss, many experiments implement purely lossy systems that are equivalent to PT-symmetric systems up to an overall exponential decay. This is commonly referred to as passive PT-symmetry. The first experimental realisations of passively PT-symmetric {\it quantum systems} have recently been achieved \cite{Xiao17,Li19,Klau19}. 

Cold atoms in optical lattices provide a natural platform for investigating PT-symmetry in many-body systems. Here we study the interplay of Bloch oscillations with PT-symmetric particle losses in a Bose-Hubbard chain. Specifically, we consider the effects of localised single-particle losses from the odd sites of the lattice, which can be described by the Lindblad equation with Lindblad operators of the form $\hat{L}_j \propto \hat{a}_j$ \cite{Sina98,Dale14}. We assume that the decay rate $\gamma$ is the same on each dissipative lattice site and take $\hat{L}_j=\sqrt{4\gamma}\hat{a}_j$, where the factor of four is included to simplify subsequent expressions. The Lindblad equation then has the form
\begin{equation}
\label{eqn:Lind}
\frac{d}{dt}\hat\rho = -i\left(\hat{H}_{\textnormal{eff}} \hat{\rho} - \hat{\rho} \hat{H}_{\textnormal{eff}}^\dag\right) - 2\gamma\sum_{j=-L}^L \left((-1)^j-1\right) \hat{a}_j \hat{\rho} \hat{a}^\dag_j,
\end{equation}
where the effective Hamiltonian is defined as
\begin{equation}
\label{eqn:Heff}
\hat{H}_{\textnormal{eff}} = \hat{H} + i\gamma\sum_{j=-L}^L \left((-1)^j-1\right) \hat{a}^\dag_j \hat{a}_j
\end{equation}
with $\hat H$ being the Bose-Hubbard Hamiltonian (\ref{eqn:BH Hamiltonian}). This is a natural first example to consider amongst many possible loss patterns. The effective Hamiltonian has passive PT-symmetry, where the parity operator interchanges even and odd sites and the time-reversal operator is simply the complex conjugation operator.  Further, the loss pattern itself is periodic, and the Bloch bands and nonlinear Bloch bands can be derived with relative ease in the single-particle and many-particle mean-field cases respectively. Some aspects of a single-particle version of this model have been investigated in \cite{Long09,Turk16,Xu16b,Long19c}. In the mean-field limit of many particles this model leads to a complex discrete nonlinear Schr\"odinger equation. Nonlinear PT-symmetric systems can have properties quite unlike their linear counterparts. They appear naturally in models of BECs in PT-symmetric potentials \cite{Grae12b,Heis13,Dast13}, as well as in the context of classical wave systems \cite{Cava11} and nonlinear optics \cite{Muss08,Rame10}. The model arising here could also be experimentally realised in coupled wave guides with absorption and Kerr nonlinearity.

In what follows we shall analyse the combined effect of particle losses and interaction/nonlinearity on Bloch oscillations. We start by considering a single-particle system in Section \ref{sec-sp}. In this case the time evolution of the state $\hat\rho$ is completely determined by $\hat{H}_\textnormal{eff}$. The spectral properties of $\hat{H}_\textnormal{eff}$ are analysed for an infinite lattice and features in the spectrum are linked to observable dynamical effects. In Section \ref{sec_mf} we consider the mean-field limit of a many-particle system, and derive analytic expressions for the generalised nonlinear stationary states and the nonlinear Bloch bands. The interplay of nonlinearity and particle losses is found to lead to peculiar features in the nonlinear Bloch bands, such as the vanishing of solutions and the formation of exceptional points. The stability of the stationary states is determined via the Bogoliubov-de Gennes equation and connected to features in the mean-field dynamics. Surprisingly, the stability of the nonlinear Bloch bands appears to influence the quantum dynamics even far from the mean-field limit. This is demonstrated numerically for a two-particle system. We end with a brief summary in Section \ref{sec_sum}.

\section{Single-particle system}
\label{sec-sp}
\subsection{State evolution}
Let us first consider the time evolution of the quantum state $\hat\rho$ for a single particle. In this case the time evolution of $\hat \rho$ is completely determined by the effective Hamiltonian (\ref{eqn:Heff}). To see this, consider the expansion of the density operator in the basis $\{|j\ra, |v\ra\}$, 
\begin{equation}
\hat \rho = \sum_{j,k} \rho_{jk} |j\ra\la k| + \sum_j \rho_{jv}|j\ra\la v| + \sum_j \bar{\rho}_{jv}|v\ra\la j| + \rho_{vv} |v\ra\la v|,
\end{equation}
where $|j\ra$ is a Fock state corresponding to a single particle on the $j$\textsuperscript{th} lattice site and $|v\ra$ is the vacuum state. Inserting  this expansion into the Lindblad equation (\ref{eqn:Lind}) yields the equations of motion for the matrix elements
\begin{equation}
i\dot{\rho}_{jk} = \left[\hat{H}_\textnormal{eff} \hat \rho - \hat \rho \hat{H}^\dag_\textnormal{eff}\right]_{jk}, \quad i\dot{\rho}_{jv} = \left[\hat{H}_\textnormal{eff}\hat{\rho}\right]_{jv}, \quad i\dot{\rho}_{vv} = -2i\gamma\sum_j \left((-1)^j-1\right)\rho_{jj}.
\end{equation}
By defining the components of the vector $\psi$ via $\rho_{jk} = \psi_j \bar{\psi}_k$ it follows from the first equation above that the evolution of $\psi$ is governed by the non-Hermitian Schr\"odinger equation
\begin{equation}
\label{eqn:Heff Schro}
i\dot{\psi} = \hat H_{\textnormal{eff}}\, \psi,
\end{equation}
where $\hat H_\textnormal{eff}$ is the single-particle effective Hamiltonian (\ref{eqn:Heff})
\begin{equation}
\label{eqn:SP Hamiltonian inf}
\hat H_{\rm eff} = \sum_{j=-L}^{L}-\left(|j+1\rangle \langle j| + |j\rangle \langle j+1|\right)  + F j |j\rangle\langle j|+i\gamma \left((-1)^j-1\right)|j \rangle\langle j|.
\end{equation}
 
The non-Hermitian Schr\"odinger equation (\ref{eqn:Heff Schro}) yields the equations of motion for the components as
\begin{equation}
\label{eqn:Psi eom}
i \dot \psi_j = -\left(\psi_{j+1} + \psi_{j-1}\right) + F j \psi_j + i\gamma\left((-1)^j-1\right)\psi_j.
\end{equation}
For a single-particle state we have $\rho_{jv}(0) = 0$ and thus $\rho_{jv}(t) = 0$ for all time. Therefore, the time evolution of $\hat\rho$ is completely determined by $\psi$,
\begin{equation}
\rho_{ij}(t) = \psi_i(t) \bar{\psi}_j(t), \quad \rho_{vv}(t) = 1 - |\psi(t)|^2,
\end{equation}
where the expression for $\rho_{vv}(t)$ follows from the trace conservation of the Lindblad equation.

\subsection{Spectral features}

Let us now examine the spectrum of the effective Hamiltonian (\ref{eqn:SP Hamiltonian inf}) with an infinite lattice. We will use algebraic methods to show that the spectrum generally consists of two ladders of eigenvalues. The analysis in this section closely follows \cite{Brei06}, in which a Hermitian one-dimensional Bose-Hubbard Hamiltonian with a variation of the on-site energy of every other site is analysed. 

Consider the complex conjugation operator $\hat T$ and the translation operator $\hat{S}_m = \sum_j |j-m\ra\la j|$, which shifts the lattice by $m$ sites to the left. A short calculation shows that this pair of operators and the Hamiltonian satisfy the commutation relations
\begin{align}
\left[\hat{S}_m,\hat{H}_\textnormal{eff}\right] &= i\gamma \sum_{j=-\infty}^{+\infty} (-1)^j \left(1-\left(-1\right)^m\right) |j-m\ra\la j| + mF  \sum_{j=-\infty}^{+\infty} |j-m\ra\la j|,\label{eqn:T H comm}\\
\left[\hat T, \hat{H}_\textnormal{eff}\right] &= -2i\gamma \sum_{j=-\infty}^{+\infty} \left((-1)^j-1\right) |j\ra\la j|\hat T,\label{eqn:C H comm}\\
\left[\hat{S}_m,\hat T\right] &= 0. \label{eqn:T C comm}
\end{align} 
Let $|\lambda\ra$ be an eigenvector of $\hat{H}_\textnormal{eff}$ with the eigenvalue $\lambda\left(F,\gamma\right)$, that is,
\begin{equation}
\label{eqn:heff eigen}
\hat{H}_\textnormal{eff} |\lambda\ra = \lambda\left(F,\gamma\right) |\lambda\ra.
\end{equation}
To simplify the notation we temporarily drop the functional dependence of $\lambda(F,\gamma)$ on the system parameters. Using the commutation relations above it can be shown that 
\begin{align}
\hat{H}_\textnormal{eff}\left[\hat{S}_{2l}|\lambda\ra\right] &= \big(\lambda - 2lF\big)\left[\hat{S}_{2l}|\lambda\ra\right],\nonumber\\
\hat{H}_\textnormal{eff}\left[\hat{S}_{2l+1}\hat T|\lambda\ra\right] &= \bigg(\bar{\lambda} - (2l+1)F - 2i\gamma\bigg)\left[\hat{S}_{2l+1}\hat T |\lambda\ra\right],
\end{align}
where $l \in \mathbb{Z}$. An application of the operators $\hat{S}_{2l}$ and $\hat{S}_{2l+1}\hat T$ to the eigenvector $|\lambda\ra$ then yields the ladders of eigenvectors
\begin{equation}
\label{eqn:heff ladder}
\hat{H}_\textnormal{eff} |E_{\alpha,n}\ra = E_{\alpha,n}|E_{\alpha,n}\ra,
\end{equation}
where $\alpha = 0,1$ labels the ladder, and the states
\begin{equation}
|E_{0,n}\ra = \hat{S}_{-2n}|\lambda\ra, \quad |E_{1,n}\ra = \hat T \hat{S}_{-(2n+1)}|\lambda\ra
\end{equation}
are associated with the energies
\begin{align}
E_{0,n} &= \Re \lambda + 2nF + i\Im \lambda,\label{eqn:ladder energy 0}\\
E_{1,n} &= \Re \lambda + (2n+1)F - i\Im \lambda - 2i\gamma. \label{eqn:ladder energy 1}
\end{align}
The expressions for these energies can be simplified by noting that the effective Hamiltonian possesses a chiral symmetry
\begin{equation}
\hat{X} \hat{H}_\textnormal{eff}^\dag \hat{X}^{-1} = -\hat{H}_\textnormal{eff},
\end{equation}
where $\hat X$ is a unitary and Hermitian operator defined as
\begin{equation}
\label{eqn:X op}
\hat{X} = \sum_j (-1)^j |-j\ra\la j|.
\end{equation}
Chiral symmetries in non-Hermitian systems have attracted considerable attention recently, see, e.g., \cite{Scho13,Malz15,Lieu18,Rive19}. The chiral symmetry of $\hat{H}_\textnormal{eff}$, together with the (passive) PT-symmetry, implies that the eigenvalues $\lambda$ have real parts symmetric around $\Re \lambda = 0$, and complex conjugate imaginary parts (up to an overall imaginary energy shift). By acting on the eigenvalue equation (\ref{eqn:heff ladder}) with $\hat{X}$ one can show that $\hat{X}|E_{\alpha,n}\ra$ is an eigenvector of $\hat{H}^\dag_\textnormal{eff}$ with eigenvalue $-E_{\alpha,n}$ and the spectrum of $\hat{H}^\dag_\textnormal{eff}$ may be written as
\begin{equation}
\label{eqn:heff dag spec1}
\sigma = \left\{-\Re \lambda - i\Im \lambda, -\Re \lambda + i \Im \lambda + 2i\gamma - F\right\}
\end{equation}
modulo $2 F$.  However, the eigenvalues of $\hat{H}^\dag_\textnormal{eff}$ can also be put into correspondence with the complex conjugate eigenvalues of $\hat{H}_\textnormal{eff}$, that is,
\begin{equation}
\label{eqn:heff dag spec2}
\sigma = \left\{\Re \lambda - i\Im \lambda, \Re \lambda + i \Im \lambda + 2i\gamma + F\right\}
\end{equation}
modulo $2 F$. Equality of the sets (\ref{eqn:heff dag spec1}) and (\ref{eqn:heff dag spec2}) implies that $\Re \lambda = 0$, and the energies in (\ref{eqn:ladder energy 0}) and (\ref{eqn:ladder energy 1}) simplify to
\begin{align}
E_{0,n} &= 2nF + i\Im \lambda\left(F,\gamma\right),\label{eqn:ladder energy 00}\\
E_{1,n} &= (2n+1)F - i\Im \lambda\left(F,\gamma\right) - 2i\gamma. \label{eqn:ladder energy 10}
\end{align}

\begin{figure}[htb]
 
        \centering
             \includegraphics[width=0.49\textwidth]{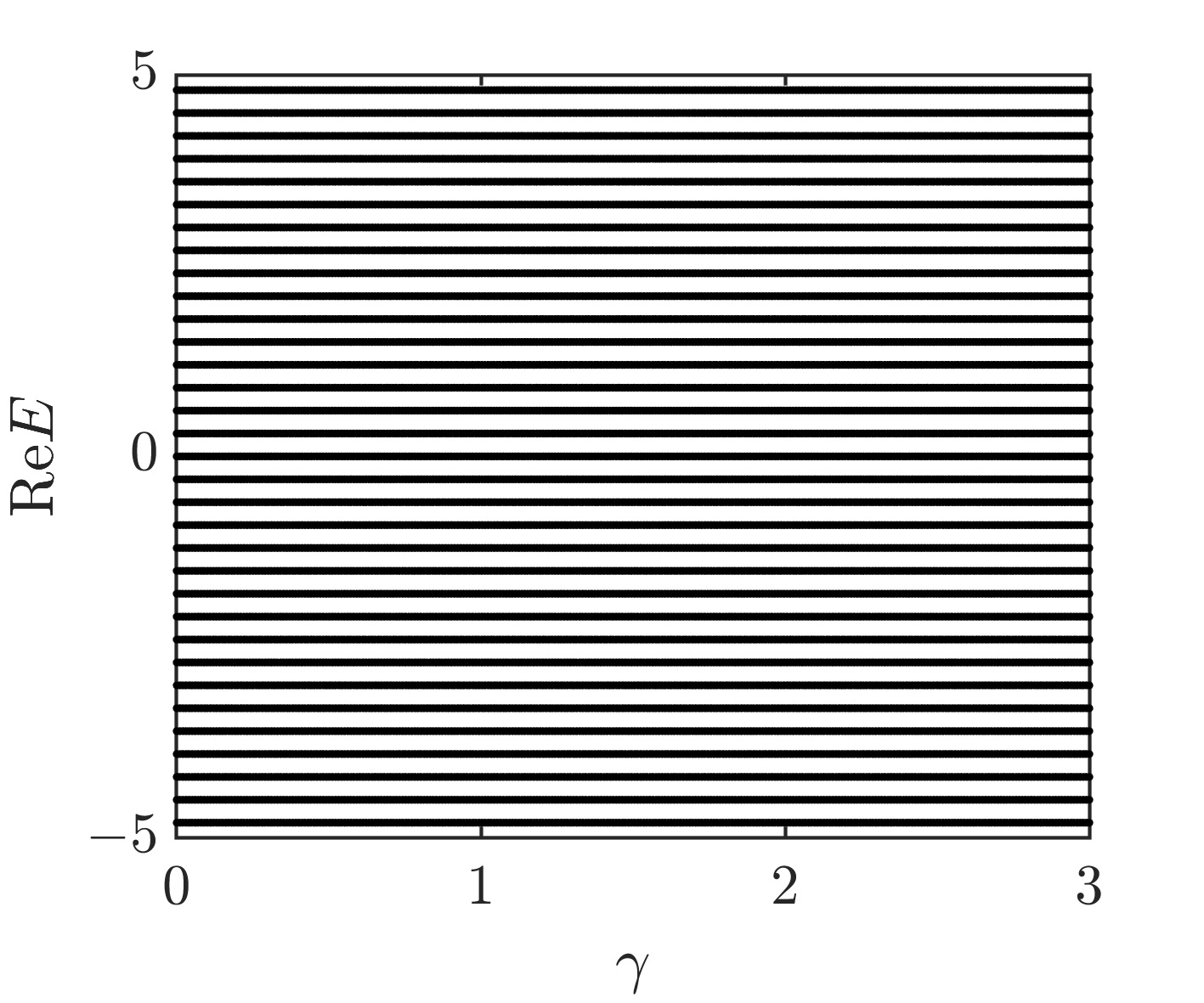}
             \includegraphics[width=0.49\textwidth]{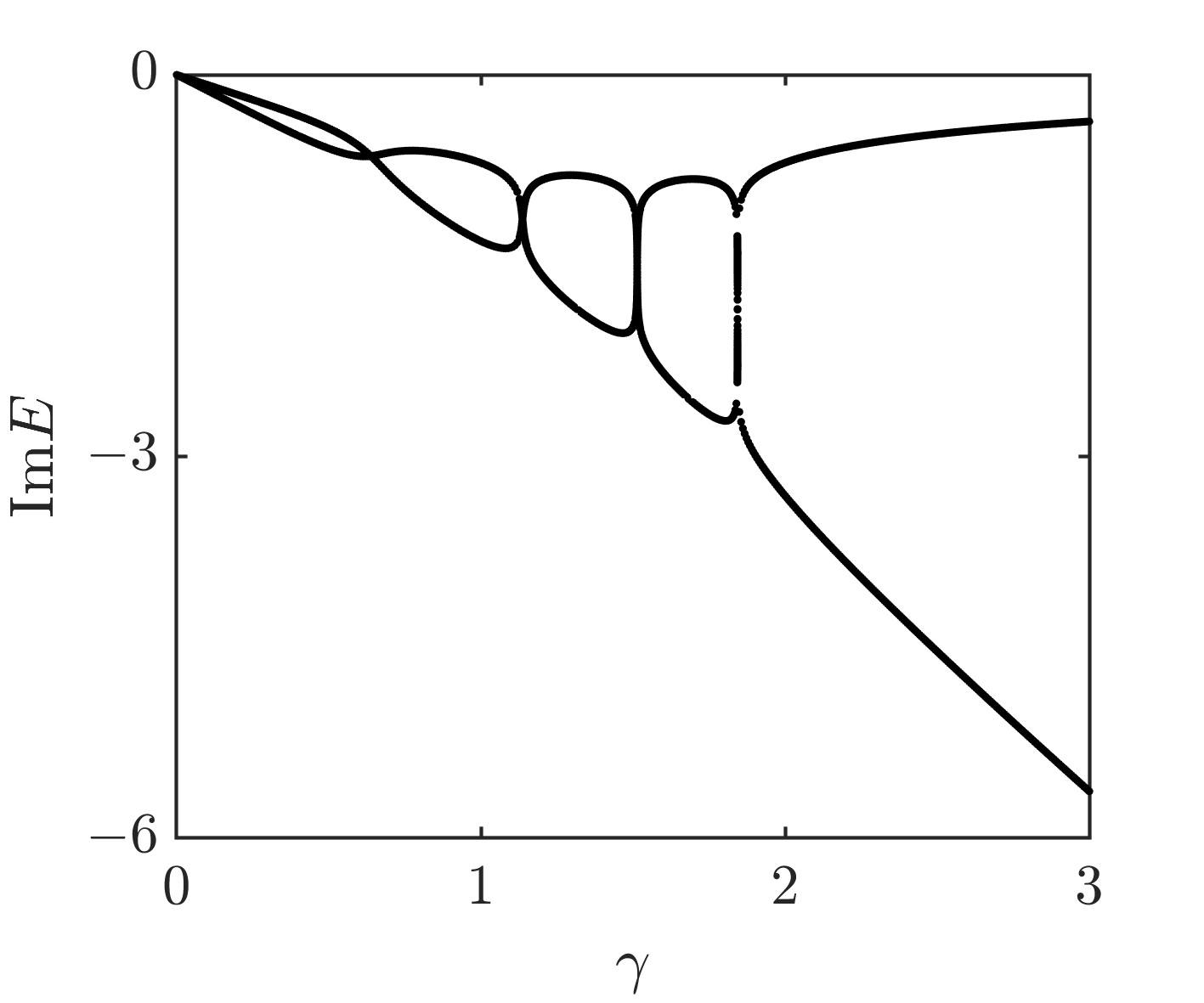}\hfill
             \includegraphics[width=0.49\textwidth]{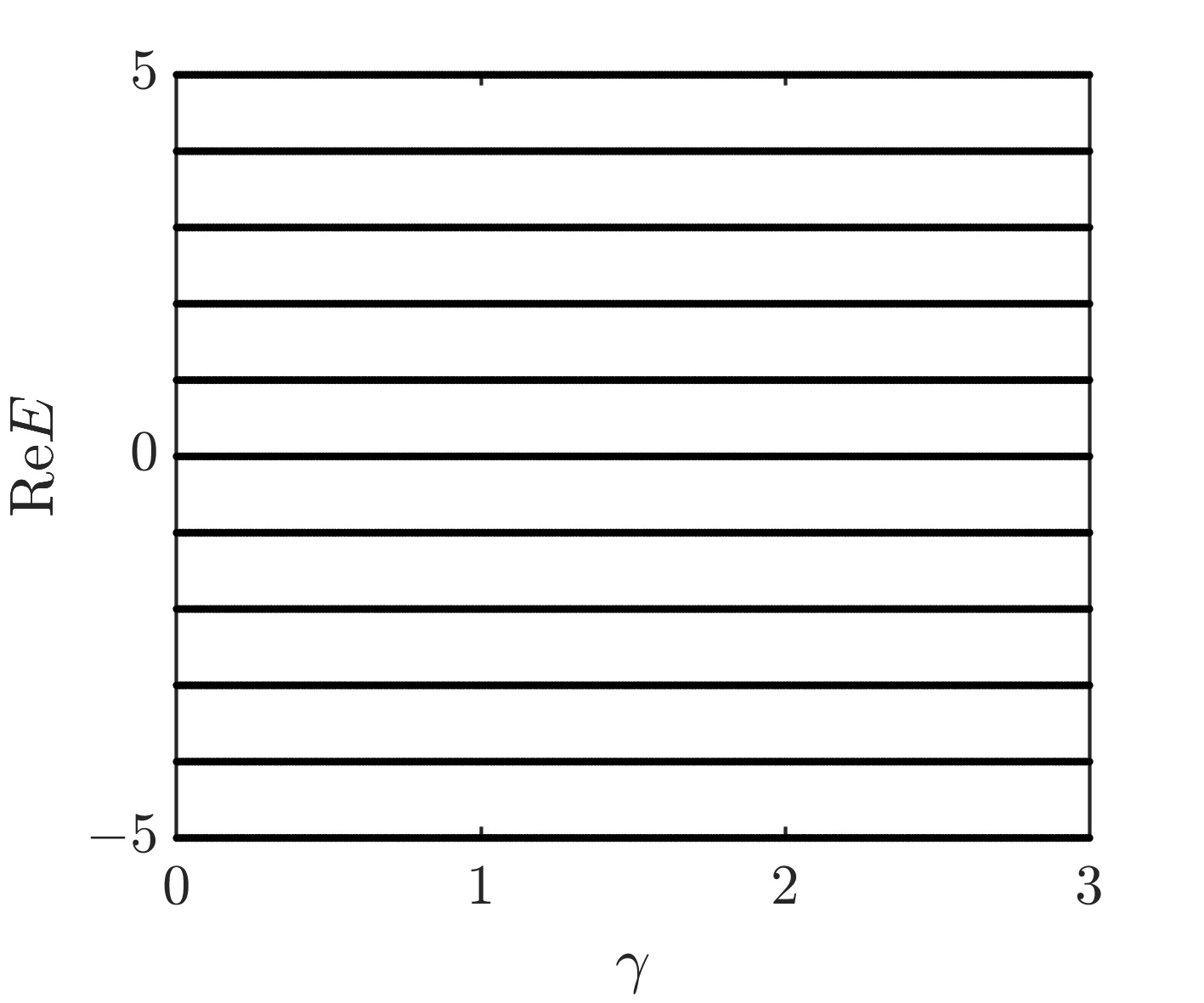}
             \includegraphics[width=0.49\textwidth]{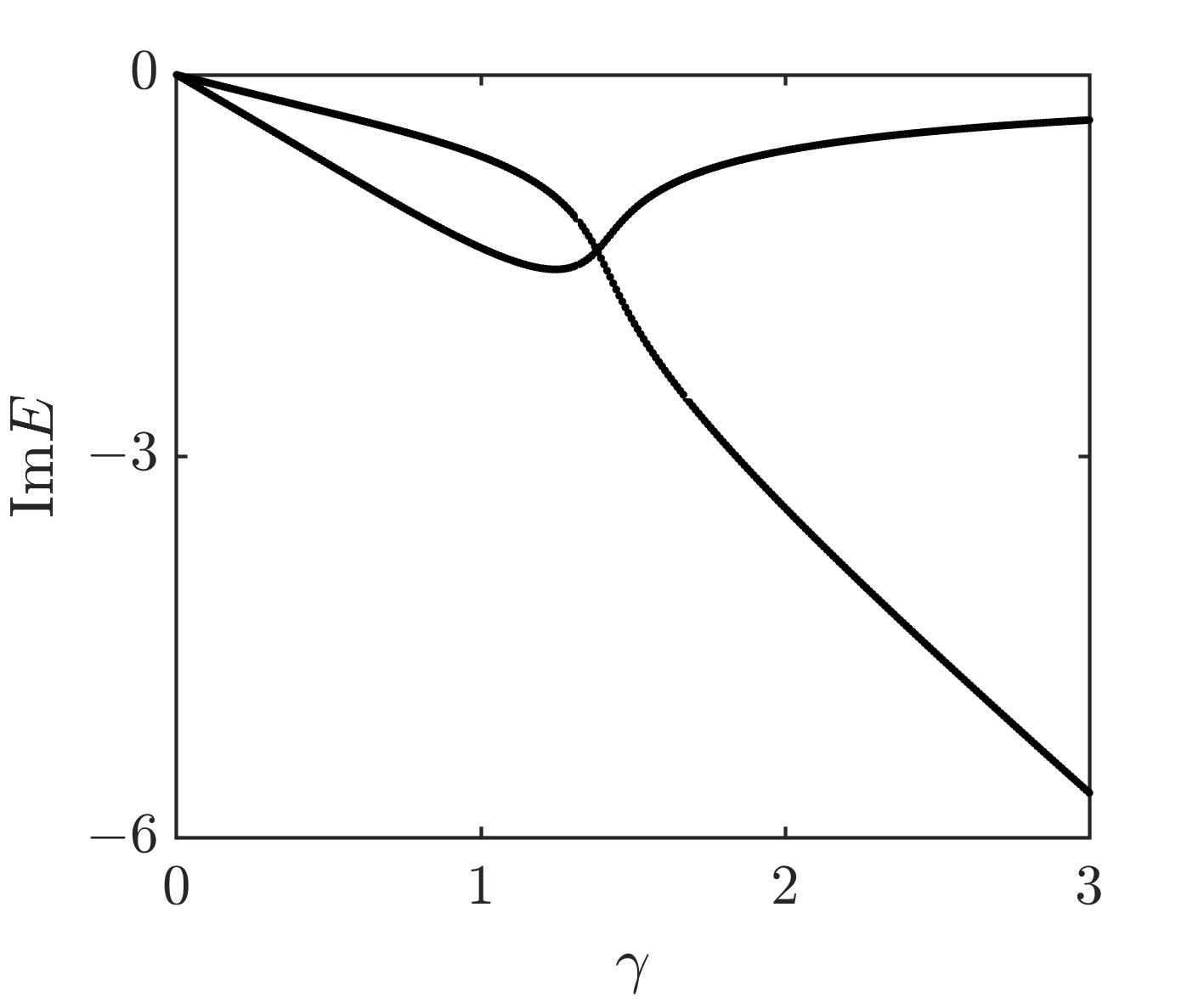}
             
             \caption{Real (left) and imaginary (right) parts of the eigenvalues of the effective Hamiltonian as a function of the decay rate $\gamma$ for a fixed static tilt $F = 0.3$ (top row) and $F = 1$ (bottom row). Obtained from numerical simulations, where eigenvalues appearing due to the finite lattice size were discarded.} 
        \label{fig:spec Heff}
\end{figure}

Thus, the spectrum of $\hat{H}_\textnormal{eff}$ generally consists of two energy ladders. The real part has values centred around $\Re E = 0$, with equal spacings of size $F$, while the imaginary part of each ladder is fixed and independent of $n$. The real and imaginary parts of the spectrum are shown in Figure \ref{fig:spec Heff} in dependence on $\gamma$ for two different values of $F$. The eigenvalues were obtained from numerical simulations and eigenvalues due to the finite lattice size were discarded. This was achieved by increasing the size of the lattice and only keeping the eigenvalues that converged with increasing lattice size.

\begin{figure}[htb]
 
        \centering
             \includegraphics[width=0.328\textwidth]{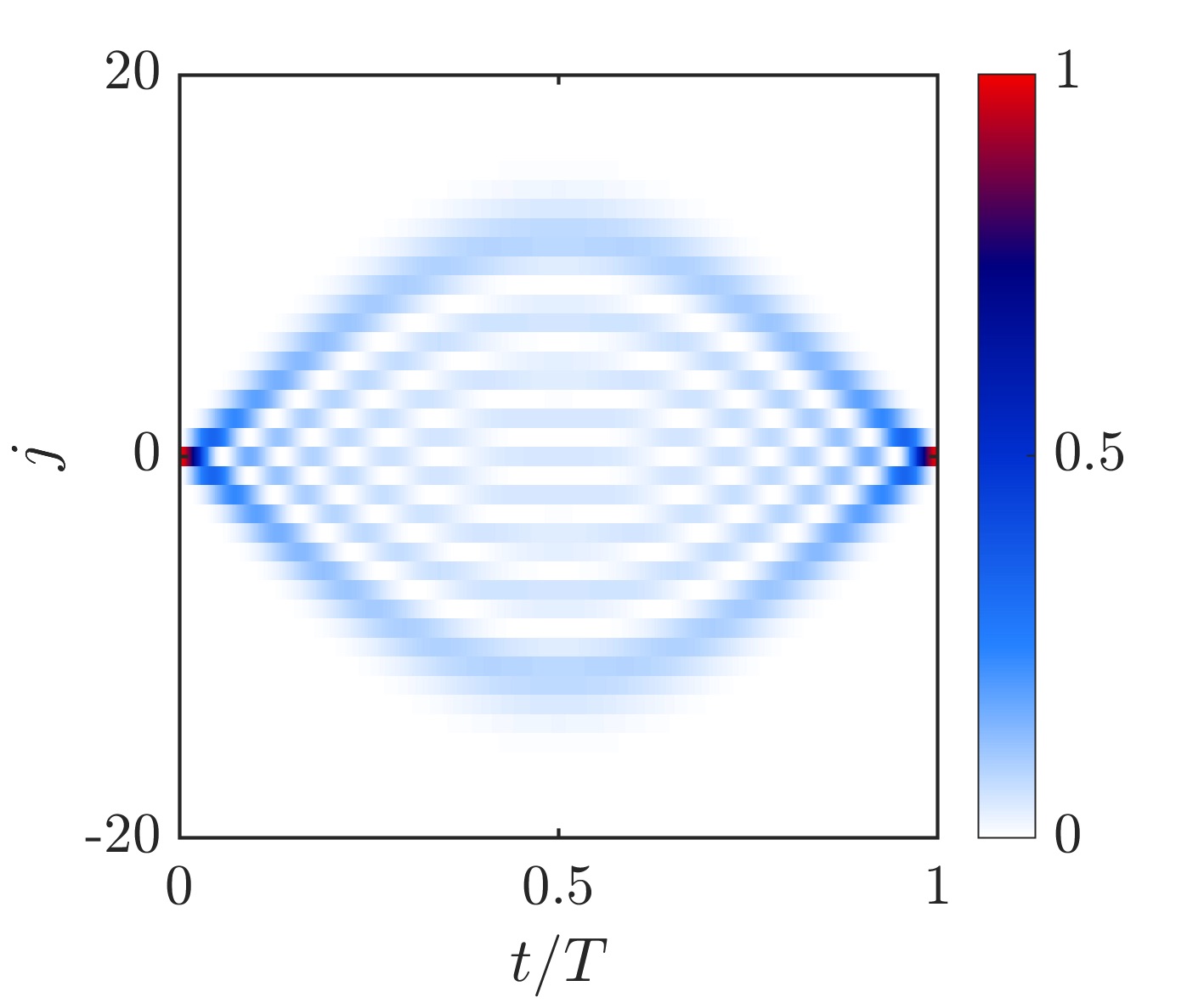}
             \includegraphics[width=0.328\textwidth]{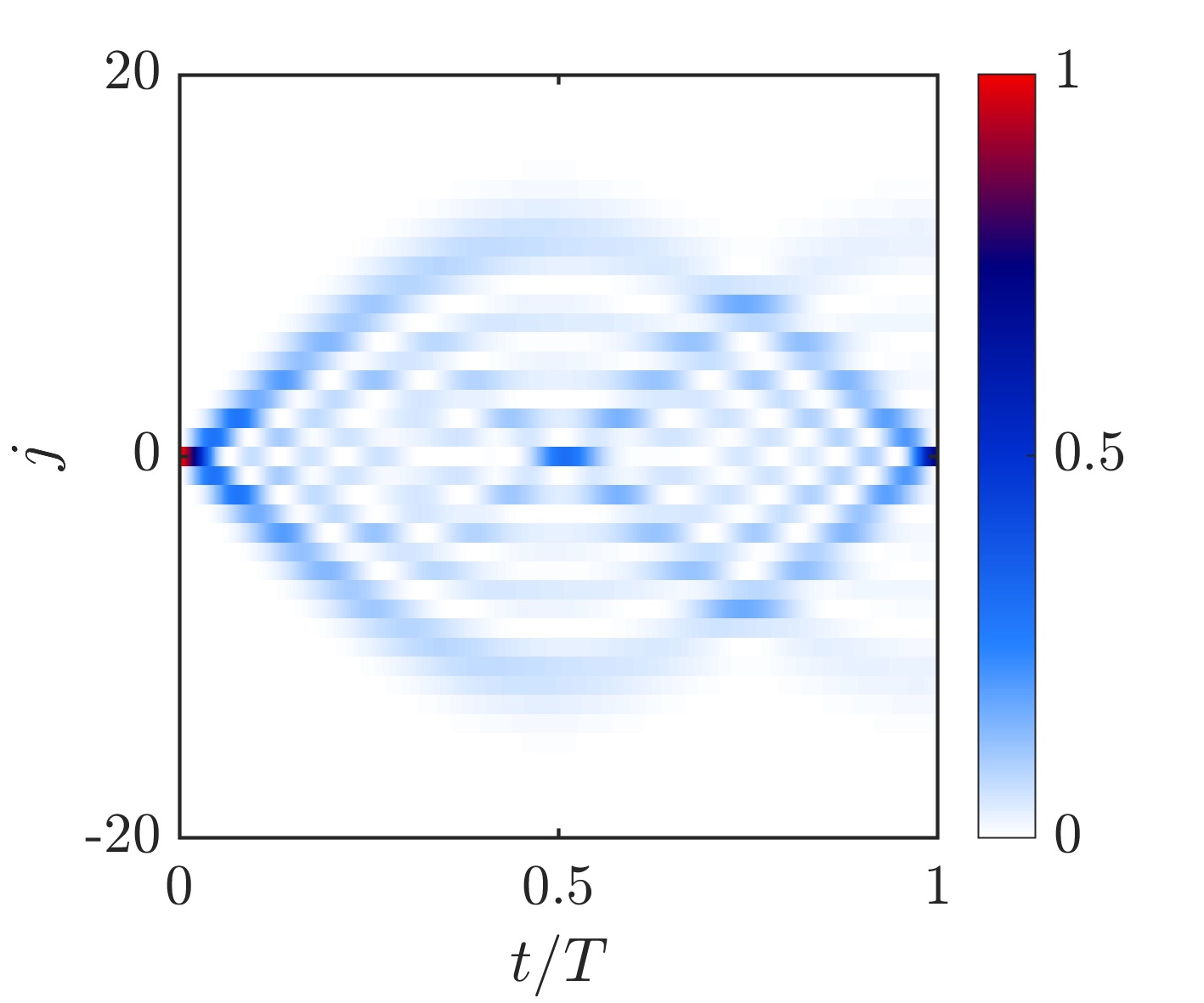}
             \includegraphics[width=0.328\textwidth]{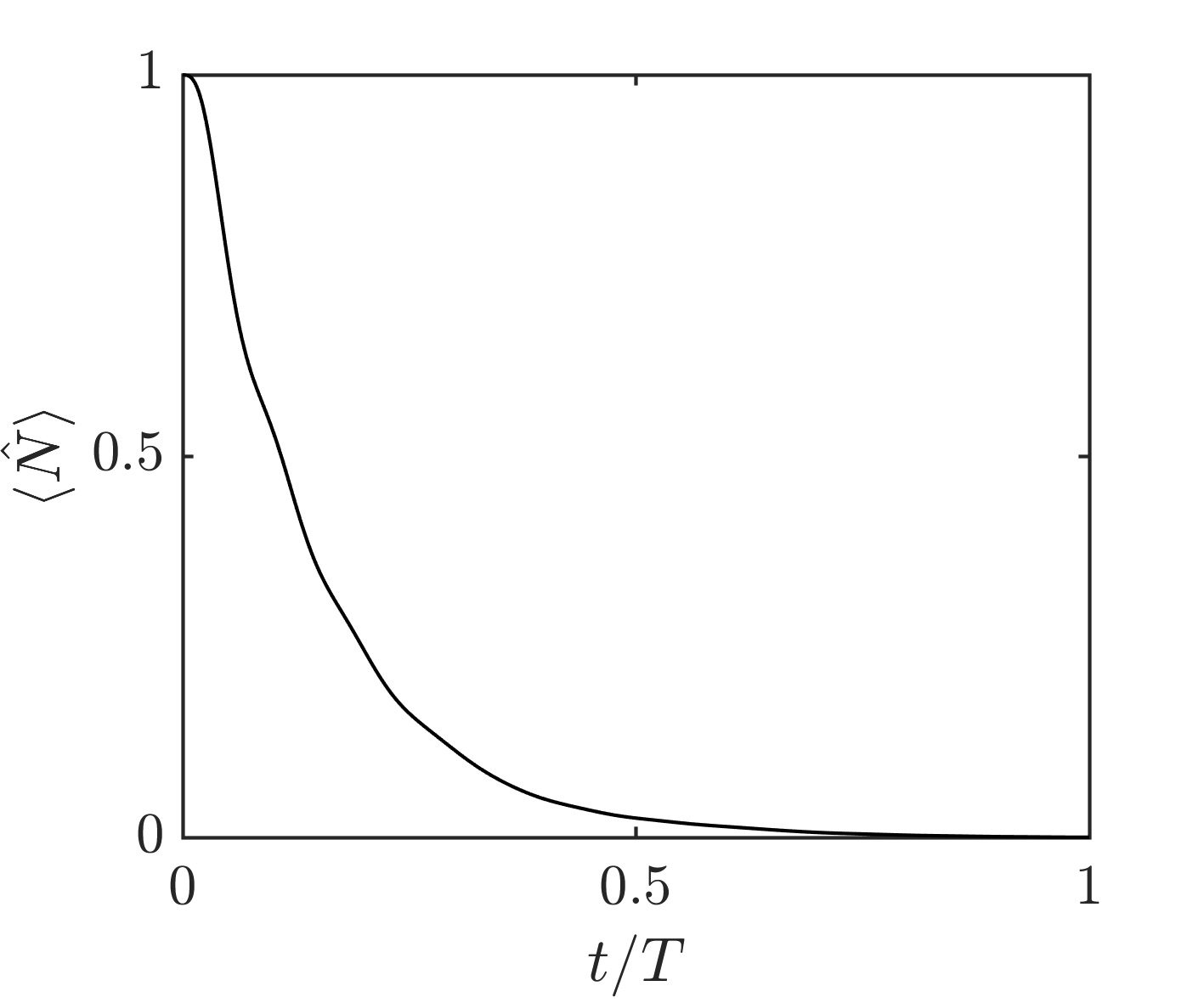}\\
              \includegraphics[width=0.328\textwidth]{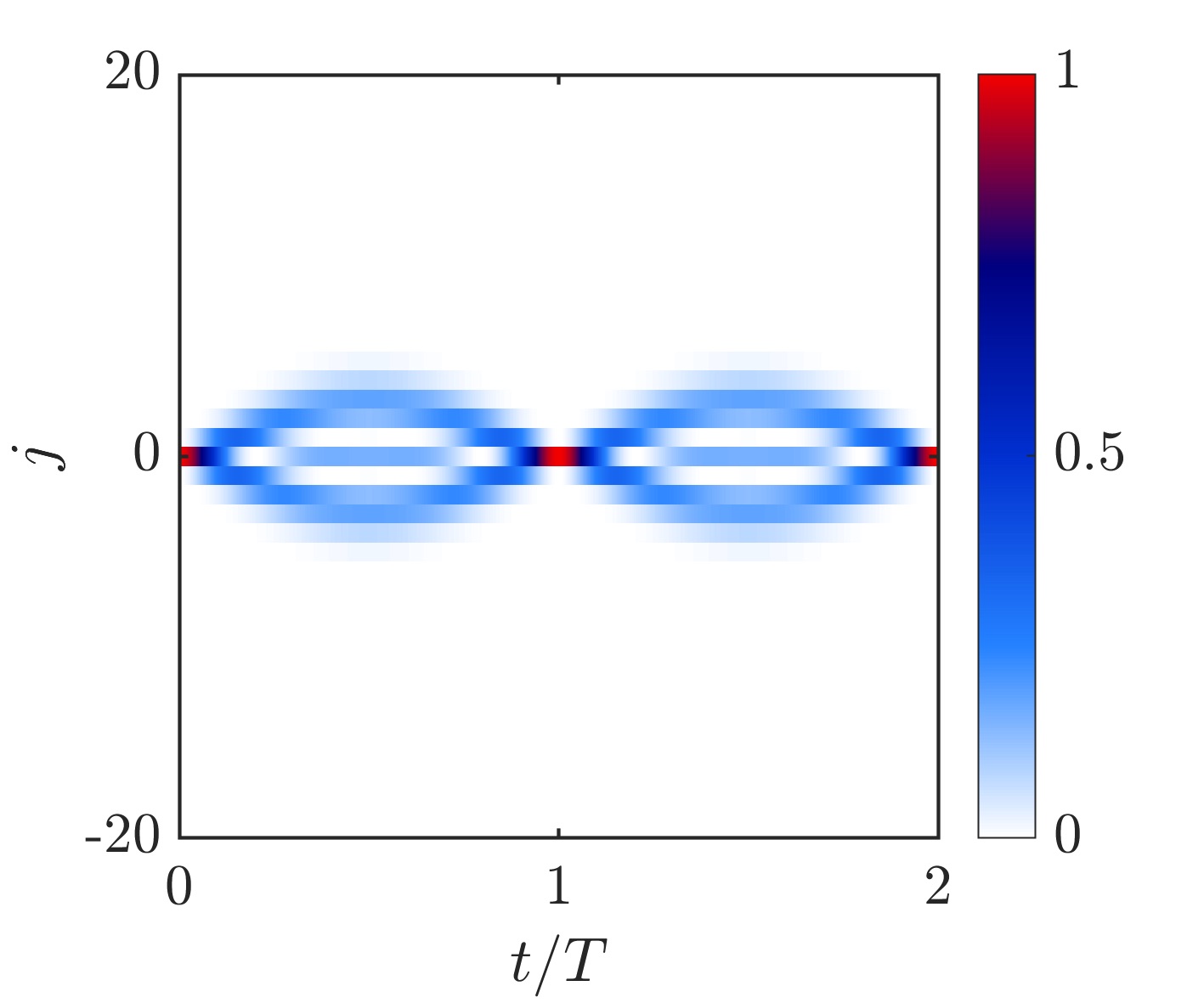}
             \includegraphics[width=0.328\textwidth]{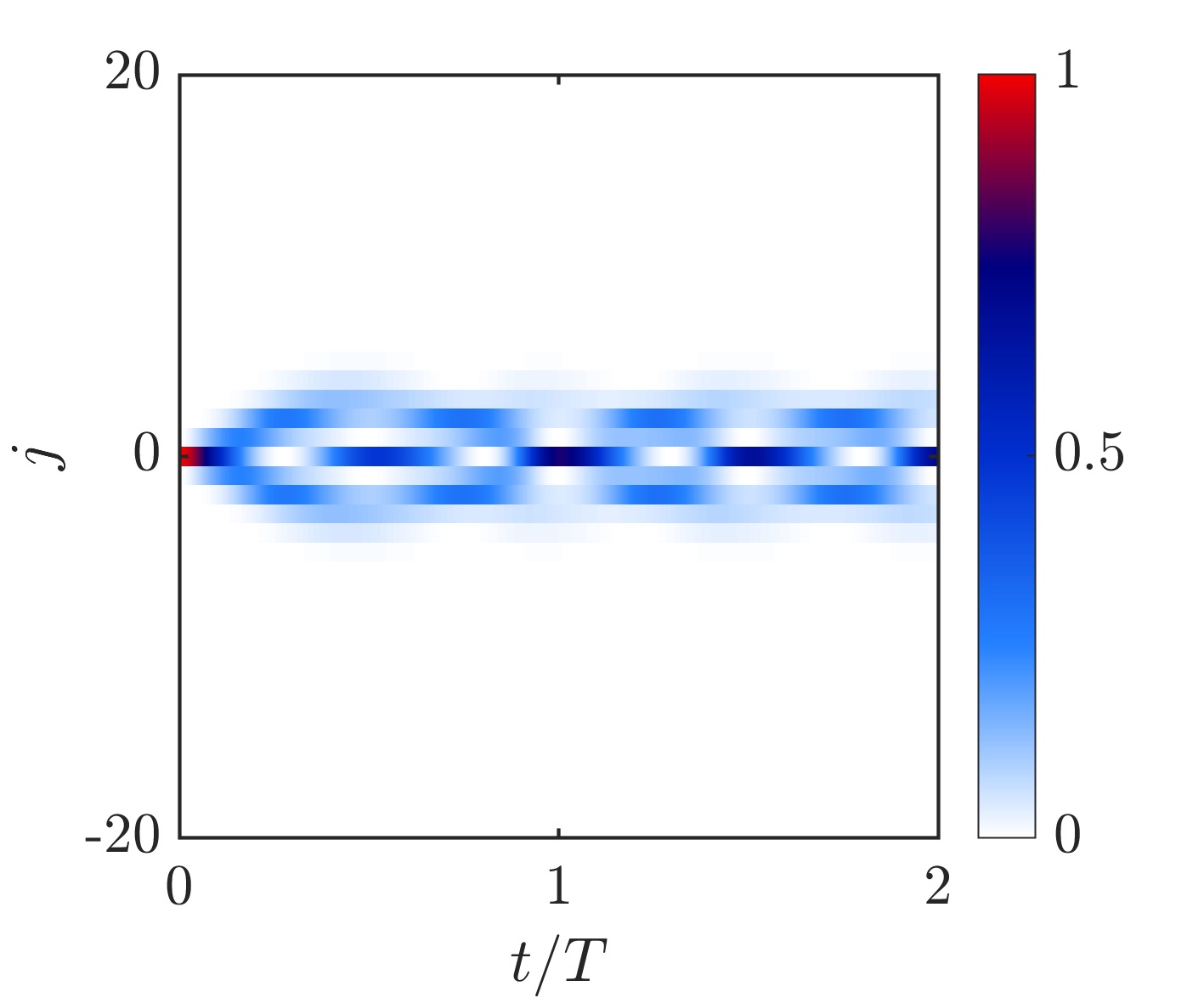}
             \includegraphics[width=0.328\textwidth]{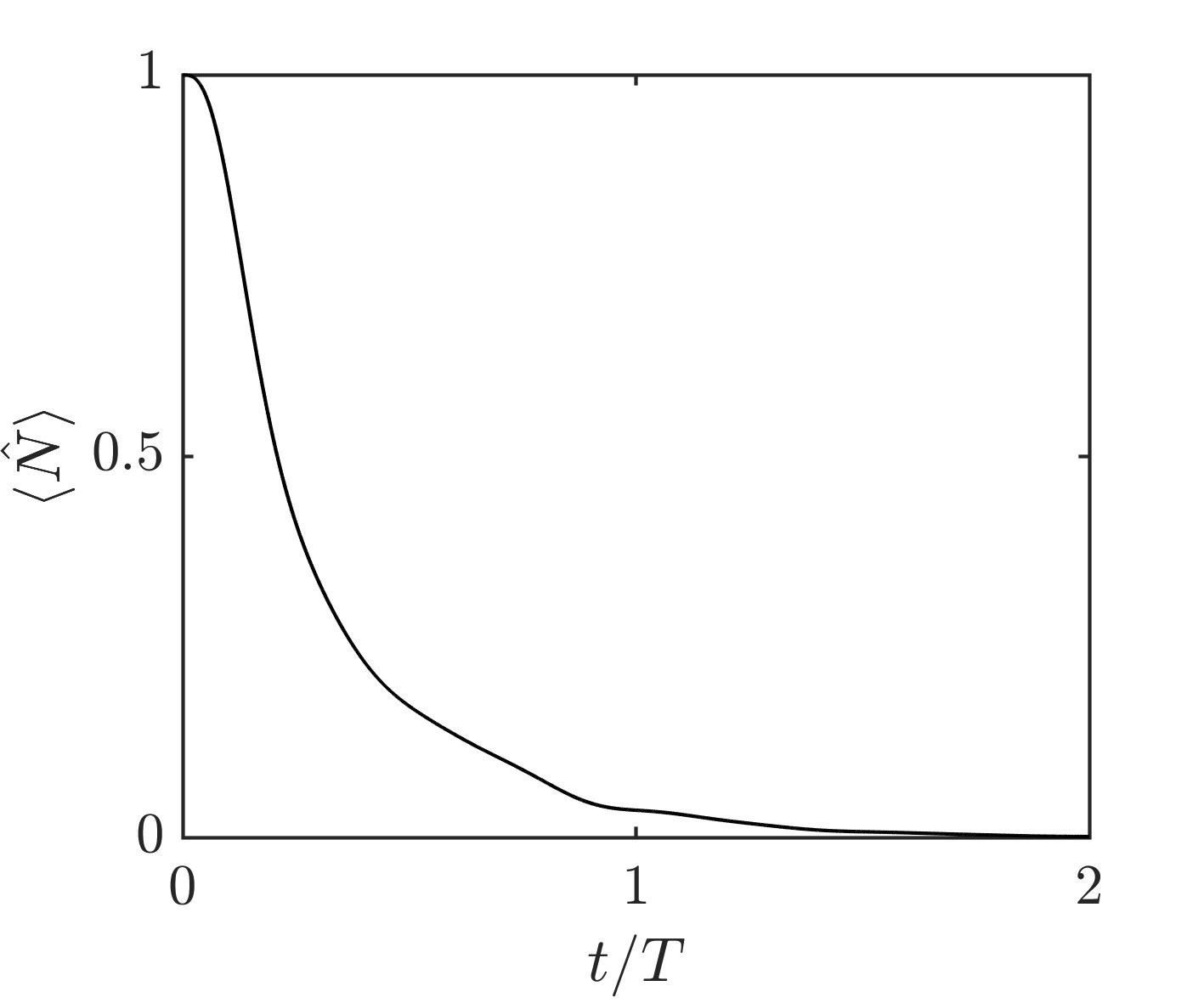}
             \caption{Renormalised density as a function of time for a single particle initialised on the central lattice site $j=0$, with a static tilt $F = 0.3$ (top row) and $F=1$ (bottom row). The unitary dynamics $\gamma = 0$ are plotted in the left column. The dissipative dynamics with $\gamma = 0.2$ and $\gamma = 0.4$ are shown in the middle column of the top and bottom rows respectively. The particle number expectation value dynamics are depicted in the right column for the dissipative cases.} 
        \label{fig:periodD}
\end{figure}

In the unitary case ($\gamma = 0$) it is well known that the dynamics are periodic with Bloch period $T = 2\pi/F$ (see, e.g., \cite{Hart04}). From the above spectral analysis we see that when $\gamma \neq 0$ the differences in the real parts of the energies are now given by $2F$ within each ladder. But the ladders are stacked such that the resulting energy difference of the ordered real parts is $F$, just as in the unitary case. This gives rise to periodic dynamics with the same Bloch period $T = 2\pi/F$. However, due to the difference in the imaginary parts of the two ladders, the components from one ladder may decay faster than those from the other. This can lead to the emergence of a motion dominated by only the stable ladder, with half the period of the original motion. For example, when $\gamma = 0$ a single particle initialised on a single lattice site $|\Psi(0)\ra = |j\ra$ performs left-right symmetric Bloch oscillations with a time period $T$ \cite{Hart04}. When $\gamma \neq 0$ the decay of one of the ladders eventually leads to an effective frequency doubling of the system, resulting in an additional structure with half the period. This effect is illustrated in Figure \ref{fig:periodD}, where the renormalised density $|\psi_j|^2/|\psi|^2$ on each lattice site $j$ is shown for both the unitary and dissipative time evolution.

\subsection{Dispersion relation}

In contrast to the breathing motion of a particle initialised on a single lattice site, a broad initial beam performs spatial Bloch oscillations, an example of which is depicted in the left panel in Figure \ref{fig:sp_beam}. In order to understand this behaviour we first derive the dispersion relation for the field-free ($F=0$) effective Hamiltonian (\ref{eqn:SP Hamiltonian inf}) with an infinite lattice. 

To this end, we consider the effective Hamiltonian (\ref{eqn:SP Hamiltonian inf}) in the basis of Bloch states
\begin{equation}
|k\ra = \sum_j |j\ra\la j|k\ra = \frac{1}{\sqrt{2\pi}}\sum_j \ue^{ikj}|j\ra,
\end{equation}
where the quasimomentum $k$ is confined to the region $-\pi \leq k \leq \pi$. The Bloch states are orthogonal and normalised to the $2\pi$-periodic delta comb $\la k|k'\ra = \delta_{2\pi}(k'-k)$. The effective Hamiltonian is not diagonal in this basis and for $F=0$ one finds the equations of motion
\begin{equation}
i\dot{\psi}(k) = \left(-2\cos k  - i\gamma\right)\psi(k) + i\gamma \psi(k+\pi).
\end{equation}
Following \cite{Long10} we introduce the two-component function $\Phi = (\Phi_1,\Phi_2)$ with $\Phi_1(k) = \psi(k)$ and $\Phi_2(k) = \psi(k+\pi)$. The time evolution can then be written as the two-component Schr\"odinger equation
\begin{equation}
i\dot{\Phi} = (-2\cos k\sigma_z + i\gamma \sigma_x - i\gamma)\Phi = h(k) \Phi,
\end{equation}
where $\sigma_i$ are Pauli matrices and the Bloch Hamiltonian is defined as
\begin{equation}
\label{eqn:Bloch Ham}
h(k) = \begin{pmatrix} -2\cos k - i\gamma & i\gamma \\ i\gamma & 2\cos k -i\gamma \end{pmatrix}.
\end{equation}

\begin{figure}[htb]
        \centering
              \includegraphics[width=0.49\textwidth]{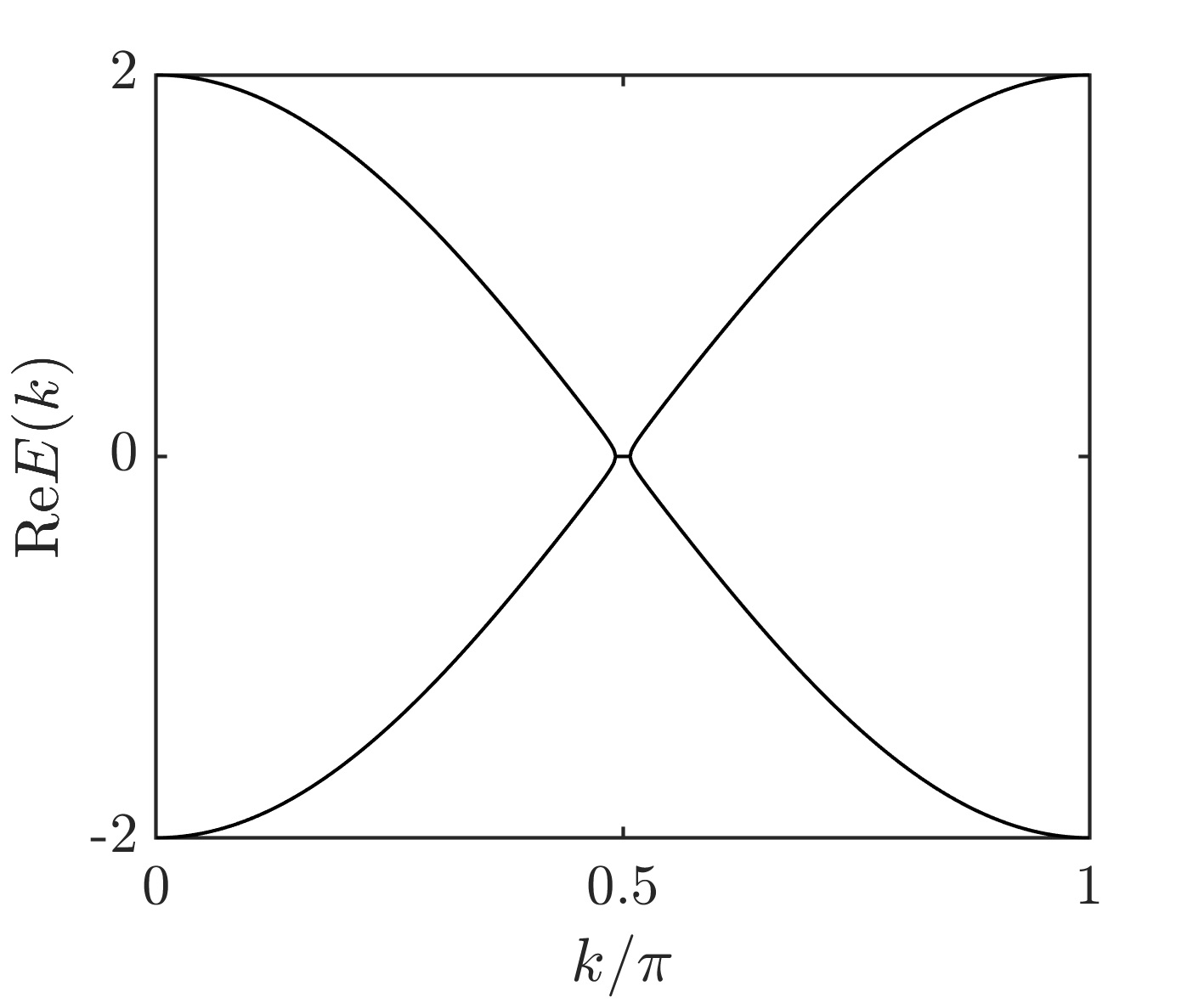}
             \includegraphics[width=0.49\textwidth]{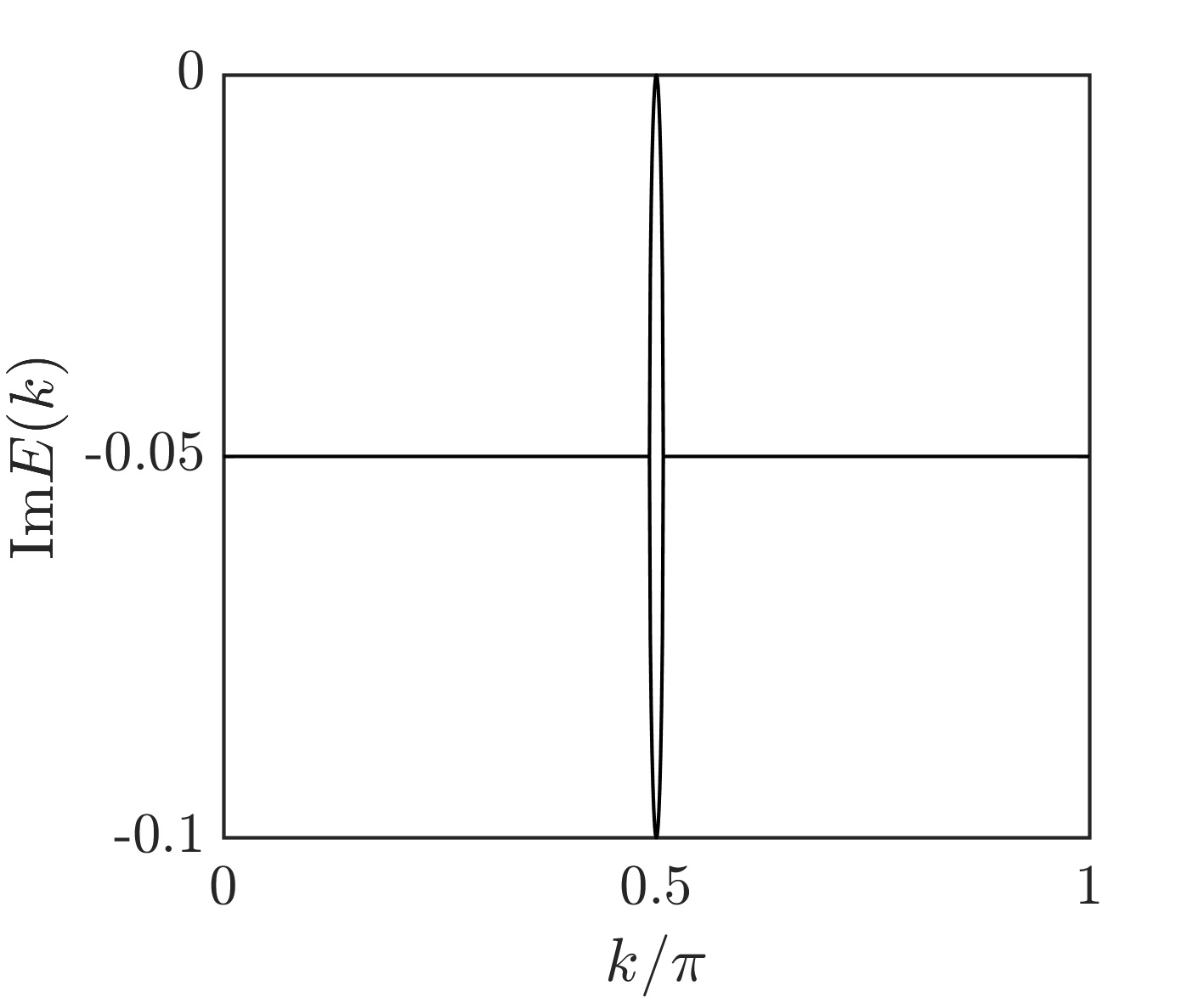}             
        \caption{Real (left) and imaginary (right) parts of the dispersion relation (\ref{eqn:dispersion}) for $\gamma = 0.05$. }
        \label{fig:spdisp}
\end{figure}

The $k$-dependent eigenvalues of $h(k)$ define the dispersion relation of the two-band system \cite{Long09,Turk16,Long19c}
 \begin{equation}
 \label{eqn:dispersion}
 E_\pm(k) = -i\gamma \pm \sqrt{4\cos^2 k - \gamma^2}.
 \end{equation}
When $\gamma< 2$ there are exceptional points at $2 |\cos k| = \gamma$, and for values of the quasimomentum $2|\cos k| > \gamma$ the energy has a real part. At the exceptional points the two eigenvalues coalesce and so do the corresponding eigenvectors. For $2|\cos k| < \gamma$ the energy becomes purely imaginary. For $\gamma\geq2$ the bands are purely imaginary for all values of the quasimomentum. The band structure is illustrated for a small decay rate of $\gamma=0.05$ in Figure \ref{fig:spdisp}.

\subsection{Beam dynamics}

\begin{figure}[htb]
        \centering
             \includegraphics[width=0.328\textwidth]{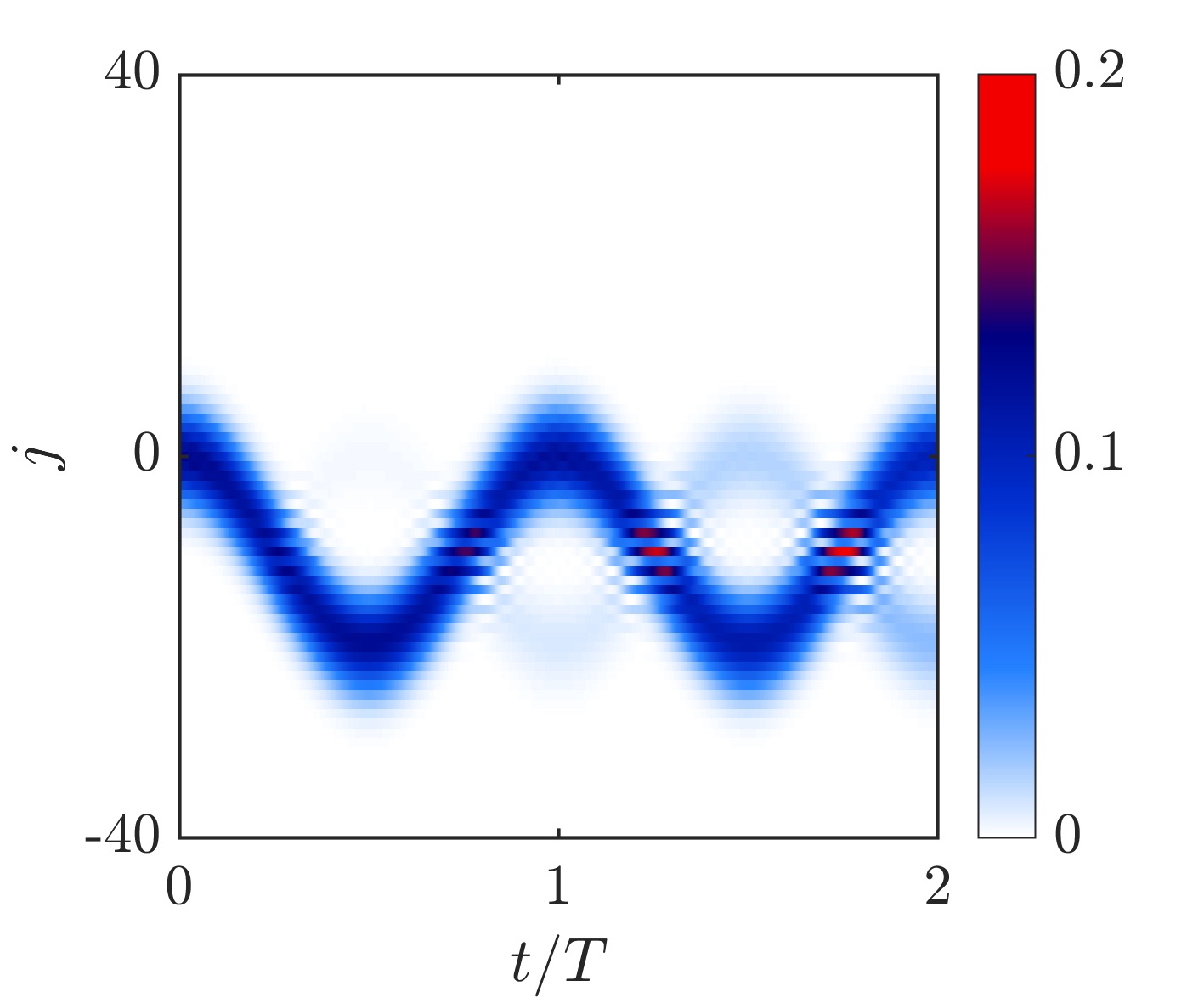}
             \includegraphics[width=0.328\textwidth]{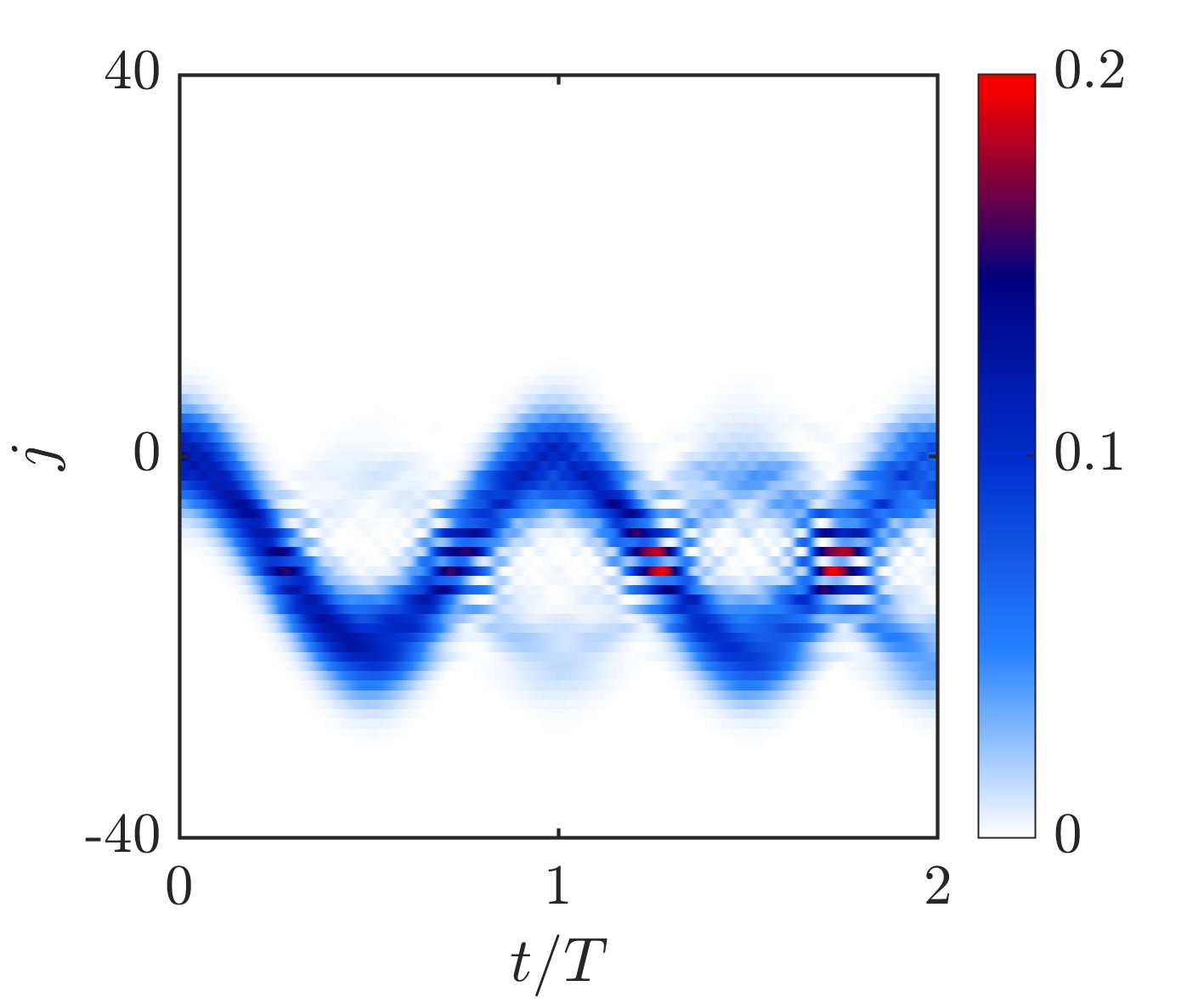}
             \includegraphics[width=0.328\textwidth]{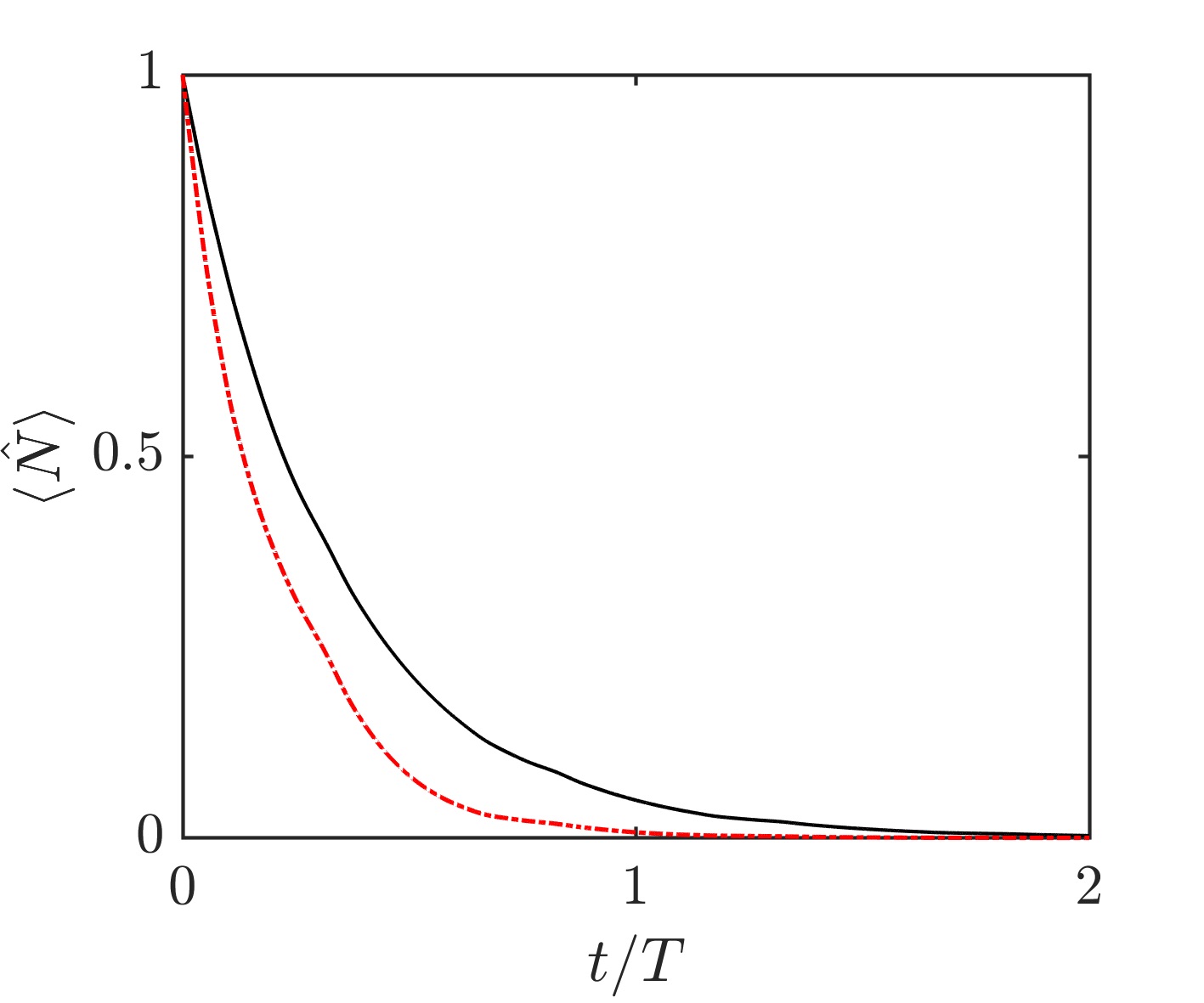}            
             \caption{The renormalised number density is plotted as a function of time for an initial Gaussian wave packet in position space with $x_0=0$, $k_0 = 0$, $\sigma = \sqrt{20}$ and $F = 0.2$. Results are shown for $\gamma = 0.05$ (left) and $\gamma_j$ randomly chosen from the interval $\left(0.025,0.125\right)$ (middle). The right frame shows the total particle number dynamics for $\gamma = 0.05$ (black solid) and the randomised decay rates (red dashed).} 
        \label{fig:sp_beam}
\end{figure}

Consider an initial Gaussian wave packet
\begin{equation}
\label{eqn:Init GB}
|\Psi(0)\ra = \mathcal{N} \sum_{j=-L}^L \ue^{-(j-x_0)^2/2\sigma^2 + ik_0(j-x_0)} |j\ra,
\end{equation}
where $x_0$ is the centre of the Gaussian; $k_0$ is the initial momentum; $\sigma$ is the width parameter; and $\mathcal{N}$ is a normalisation constant. In \cite{Long19c} it was shown that a broad Gaussian beam in position space is an approximate eigenstate of the Bloch Hamiltonian (\ref{eqn:Bloch Ham}), provided the Bloch states are close to the standard basis vectors. Furthermore, a static tilt can result in Bloch oscillations, together with a splitting of the beam in real space due to a transfer of population between the Bloch bands. This is illustrated in the left frame of Figure \ref{fig:sp_beam} for $\gamma = 0.05$ and $F=0.2$, where the Gaussian wave packet effectively maps out the real part of the band structure depicted in Figure \ref{fig:spdisp}. Numerical simulations show that this behaviour does not require the decay rate to be exactly the same on each dissipative site. This can be seen in the middle panel of Figure \ref{fig:sp_beam}, where the decay rates on the dissipative sites $\gamma_j$ are randomly selected from the interval $\left(0.025,0.125\right)$. Thus, a robust experimental realisation should be feasible. Indeed, a similar phenomenon has already been observed experimentally in PT-symmetric photonic lattices \cite{Wimm15}. 

Let us now attempt to gain some insight into the oscillatory behaviour of the beam in position space, using a quasiclassical argument similar to \cite{Hart04,Brei06,Grae16b}. In the present case there are two energy bands that degenerate at the exceptional points. In a semiclassical treatment we consider the dynamics on each of the branches of the band structure separately, and later account for transitions between the bands using a Landau-Zener type formula. 

The application of a static tilt $F$ introduces a term $Fx$ into the Bloch Hamiltonian
\begin{equation}
h(x,k) = -2\cos k \sigma_z + i\gamma \sigma_x + Fx - i\gamma,
\end{equation}
where $x = id/dk$ is canonically conjugate to $k$ with $[x,k] = i$. Note that $(x,k)$ are operators in the quasimomentum representation. The phase-space representation of $h(x,k)$ is given by the $2 \times 2$ matrix-valued function 
\begin{equation}
\label{eqn:ps dirac}
H_s(q,p) = -2\cos p \sigma_z + i\gamma \sigma_x + Fq - i\gamma,
\end{equation}
where $(q,p) \in \R \times \R$ are a pair of canonical phase-space coordinates. The eigenvalues of (\ref{eqn:ps dirac}) are readily found to be
\begin{equation}
\label{eqn:hkq sb ham}
\xi_\pm(q,p) = F q + E_{\pm}(p),
\end{equation}
where $E_\pm(p)$ is the dispersion relation (\ref{eqn:dispersion}), i.e., $E_\pm(p)$ are the eigenvalues of the field-free Bloch Hamiltonian (\ref{eqn:Bloch Ham}). In the single-band approximation the population in each band $\pm$ is assumed to evolve independently according to the semiclassical dynamics generated by the non-Hermitian Hamiltonian $\xi_{\pm}(q,p)$. Thus, as the coupling between the bands is ignored, this approximation cannot account for any population transfer between the bands. Note that the $q$ and $p$ are different in the $\pm$ bands, so that technically we should write $\xi_\pm(q_\pm,p_\pm)$. However, for notational simplicity we do not explicitly indicate this, as it should be clear from the context which band is being referred to.

The semiclassical dynamics of a Gaussian wave packet evolving under a non-Hermitian phase-space Hamiltonian $H(q,p)=\Re H+i \Im H$ are given by \cite{Grae11}
\begin{eqnarray}
\dot q&=\pa_p\Re H + \Sigma_{qp}\pa_p\Im H + \Sigma_{qq}\pa_q\Im H,\label{eqn:qband eom}\\
\dot p&=-\pa_q\Re H + \Sigma_{pp}\pa_p\Im H + \Sigma_{qp}\pa_q\Im H,\label{eqn:kband eom}
\label{eqn:semiqk}
\end{eqnarray}
where $\Sigma$ encodes the (co)variances of the canonical coordinates according to
\begin{equation}
\Sigma_{qq}=2(\Delta q)^2, \quad
\Sigma_{pp}=2(\Delta p)^2,
\quad \Sigma_{qp}=\Sigma_{pq}=2\Delta_{qp},
\end{equation}
and the determinant of $\Sigma$ is one. The covariances are also time dependent, following the dynamical equations
\begin{eqnarray}
\dot\Sigma=\Omega \Re H''\Sigma-\Sigma \Re H''\Omega + \Omega \Im H'' \Omega + \Sigma \Im H''\Sigma,\label{eqn:sigband eom}
\end{eqnarray}
where $\Omega$ is the symplectic matrix
\begin{equation}
\Omega=\begin{pmatrix} 0 &1\\-1&0\end{pmatrix}
\end{equation}
and $H''$ is the Hessian of $H$. We must distinguish between the regions where the square root $\sqrt{4 \cos^2 p - \gamma^2}$ appearing in $E_\pm(p)$ is real or imaginary. In the region where it is real we find
\begin{equation}
\label{eqn:sband real r}
\Re \xi_\pm = F q + E_\pm(p), \quad \Im \xi_\pm = -\gamma,
\end{equation}
and an application of the semiclassical equations (\ref{eqn:qband eom}) and (\ref{eqn:kband eom}) yields
\begin{equation}
\label{eqn:semiqk real}
\dot q = \pa_p \textnormal{Re}E_{\pm}(p), \quad \dot p = -F.
\end{equation}
Therefore, the acceleration theorem $p(t)=-Ft + p(0)$ is recovered and 
\begin{equation}
\label{eqn:pos real map}
q(t)=q(0) + \frac{1}{F}\left[\textnormal{Re}E_\pm(p(0)) - \textnormal{Re}E_\pm(p(0)-Ft)\right].
\end{equation}
Similarly, in the region where the square root is imaginary we have
\begin{equation}
\label{eqn:sband imag r}
\Re \xi_\pm = F q, \quad \Im \xi_\pm = -\gamma \pm \sqrt{\gamma^2-4 \cos^2 p}.
\end{equation}
The general semiclassical equations of motion then yield 
\begin{equation}
\label{eqn:semiqk imag}
\dot q = \Sigma_{qp} \pa_p \Im \xi_\pm, \quad \dot p = -F + \Sigma_{pp} \pa_p \Im \xi_\pm,
\end{equation}
and 
\begin{equation}
\dot\Sigma_{qq} = \left(\Sigma^2_{qp}-1\right)\pa_{pp}\Im \xi_\pm, \quad \dot\Sigma_{qp} = \Sigma_{qp}\Sigma_{pp}\pa_{pp}\Im \xi_\pm, \quad \dot\Sigma_{pp} = \Sigma^2_{pp} \pa_{pp} \Im \xi_\pm,
\end{equation}
where $\Im \xi_\pm$ is defined in (\ref{eqn:sband imag r}). Now observe that in the approximation of an initially infinitely narrow momentum wave packet with $\Sigma_{pp}(0)=0$, the (co)variances remain constant for all times, i.e.,  $\Sigma_{pp}(t)=0$ and $\Sigma_{qp}(t)=\Sigma_{qp}(0)$. In this case the equations of motion simplify to 
\begin{equation}
\dot q=\Sigma_{qk}(0)\pa_k \Im \xi_\pm, \quad \dot k = -F,
\end{equation}
and the acceleration theorem $p(t)=-Ft + p(0)$ is again recovered. In the special case where $\Sigma_{qp}(0)=0$ we find that $q(t) = q(0)$. The position of each wave packet follows the real dispersion relation in the unbroken passive PT-symmetry phase (\ref{eqn:pos real map}) and remains stationary in the broken phase. Therefore, as observed in Figure \ref{fig:sp_beam}, the position maps out the real part of the band structure of the Bloch Hamiltonian. 

In our approximation we have assumed that the semiclassical non-Hermitian dynamics above can be applied to each band individually. A more careful treatment would derive the semiclassical limit of non-Hermitian Dirac-type equations, thus extending the semiclassical dynamics identified in \cite{Spoh00} for the unitary case. This is an interesting problem for future investigations. Close to the exceptional point, at which the two branches of the band structure intersect, a partial transition to the other branch of the band structure takes place. While this cannot be accounted for in the single-band approximation, it can be described heuristically by a Landau-Zener-type scenario. Details of the calculation can be found in \cite{Long19c}. Here we simply quote the estimate of the transition probability between the bands, 
\begin{equation}
P=\left(2-\ue^{-\frac{\pi\gamma^2}{2F}}\right)^{-1},
\end{equation}
which fits the numerical results with surprising accuracy for a large range of parameters.

\section{Mean-field limit of many particles}\label{sec_mf}

Let us now consider the time evolution of a pure BEC, defined as
\begin{equation}
\label{eqn:bec state}
|\Psi(0)\ra = \frac{1}{\sqrt{N_0!}}\left(\sum_{j=-L}^L \psi_j \hat{a}^\dag_j\right)^{N_0} |v\ra. 
\end{equation}
Here $|v\ra$ is the vacuum state, $N_0$ is the number of particles and one can choose the coefficients $\psi_j$ to be normalised to unity,
\begin{equation}
\label{eqn:psi coeff norm}
\sum_{j=-L}^L |\psi_j|^2 = 1.
\end{equation}
A far richer set of dynamics are expected to be possible for many-particle systems, owing to the interplay between particle interactions and dissipation. However, for many particles on a large lattice the size of the Hilbert space prohibits numerical investigations. Thus, to obtain insights into the case of large particle numbers we consider the mean-field limit $N_0 \to \infty$. For unitary time evolution it is well known that the mean-field dynamics provide a good approximation, provided that the state remains close to a pure BEC for all time \cite{Mors06}. For a large number of weakly-interacting particles, the mean-field approximation is also valid for dissipative processes described by the Lindblad equation (see, e.g., \cite{Kord15}).

One way to construct the mean-field approximation is by working with the elements of the single-particle density matrix (SPDM) $\sigma_{ij} =  \la \hat{a}^\dag_i \hat{a}_j\ra/N_0$. Due to the interaction term, the equations of motion for the SPDM elements do not form a closed system of equations. Rather, they form the so-called Bogoliubov-Born-Green-Kirkwood-Yvon hierarchy. In the mean-field approximation the hierarchy is truncated by approximating the four-point correlation functions as
\begin{equation}
\label{eqn:four point}
\la \hat{a}^\dag_i \hat{a}_j \hat{a}^\dag_k \hat{a}_l \ra \approx \la \hat{a}^\dag_i \hat{a}_j \ra \la \hat{a}^\dag_k \hat{a}_l\ra.
\end{equation}
The mean-field evolution of the SPDM elements is then found to be
\begin{align}
i\dot {\sigma}_{jk} &= \left(\sigma_{j+1k} + \sigma_{j-1k} - \sigma_{jk+1} - \sigma_{jk-1}\right) + g \left(\sigma_{kk}-\sigma_{jj}\right)\sigma_{jk} \nonumber\\&- F(j-k)\sigma_{jk} + i\gamma\left((-1)^j+(-1)^k-2\right)\sigma_{jk},
\end{align}
where $g = UN_0$ is fixed as $N_0 \to \infty$. By identifying $\sigma_{jk} = \bar{\psi}_j \psi_k$ the mean-field equations may equivalently be described by the complex discrete nonlinear Schr\"{o}dinger equation
\begin{equation}
\label{eqn:cdnls tilt}
i\dot{\psi}_j = -\left(\psi_{j+1} + \psi_{j-1}\right) + F j \psi_j + g|\psi_j|^2 \psi_j + i\gamma\left((-1)^j-1\right)\psi_j.
\end{equation}
The initial conditions $\psi_j(0)$ are the parameters $\psi_j$ appearing in the pure BEC state (\ref{eqn:bec state}). The same equations of motion can also be obtained directly from a general dissipative phase-space dynamics derived from Gaussian states in \cite{Grae18}. We note that (\ref{eqn:cdnls tilt}) could also be implemented in classical optics using an array of waveguides with absorption and Kerr nonlinearity (see, e.g. \cite{Muss08,Rame10}).

\subsection{Nonlinear Bloch bands}
For a single particle we found that a knowledge of the band structure was crucial for understanding the Bloch oscillations of a broad beam. In the mean-field limit it is possible to define nonlinear Bloch states and their corresponding nonlinear generalised eigenvalues. In closed systems it has been shown that the presence of particle interactions can lead to novel features in the nonlinear Bloch bands. For instance, the formation of loops at the band edges \cite{Wu03} for a Bose-Einstein condensate in a bichromatic lattice \cite{Witt11}. Here we shall examine the nonlinear band structure when the odd lattice sites are dissipative, by deriving a nonlinear complex Bloch Hamiltonian, and following the procedure in \cite{Grae12b} to deduce its nonlinear eigenvalues.

The nonlinear Bloch states are defined as the stationary states $\phi_j$  of the field-free version of equation (\ref{eqn:cdnls tilt}) 
\begin{equation}
\mu \phi_j = -\left(\phi_{j+1} + \phi_{j-1}\right) + g|\phi_j|^2 \phi_j + i\gamma\left((-1)^j-1\right)\phi_j,
\end{equation}
where $\mu$ is a generalised nonlinear complex eigenvalue. The translational symmetry motivates the ansatz
\begin{equation}
\label{eqn:nl A and B}
\phi_j = \left\{
                \begin{array}{ll}
                  A\ue^{ikj}, \,\, j \,\,\textnormal{even},\\
                  B\ue^{ikj}, \,\, j \,\,\textnormal{odd},
                \end{array}
              \right.
\end{equation}
for the stationary states, with the quasimomentum $-\pi \leq k \leq \pi$ and the normalisation convention $|A|^2 + |B|^2 = 1$. The coefficients $A$ and $B$ are then determined by the nonlinear two-level system
\begin{equation}
\label{eqn:mf two level}
\begin{pmatrix} g |A|^2 & -2\cos k\\-2\cos k & g |B|^2 - 2i\gamma \end{pmatrix} \begin{pmatrix} A \\ B \end{pmatrix} = \mu \begin{pmatrix} A \\ B \end{pmatrix}.
\end{equation}
To simplify the calculations we subtract an overall complex energy $\frac{g}{2} - i\gamma$ from the diagonal, leading to the PT-symmetric system
\begin{equation}
\label{eqn:nonlin mat}
\begin{pmatrix} cz+i\gamma & -2\cos k\\-2\cos k & -cz -i\gamma \end{pmatrix} \begin{pmatrix} A \\ B \end{pmatrix} = \tilde\mu \begin{pmatrix} A \\ B \end{pmatrix},
\end{equation}
with
\begin{equation}
z = |A|^2 - |B|^2, \quad c = \frac{g}{2}.
\end{equation}

In \cite{Grae12b} it was shown that the eigenvalues of a nonlinear complex two-level system of the form (\ref{eqn:nonlin mat}) can be obtained analytically. Here we closely follow the procedure in \cite{Grae12b} to obtain the nonlinear Bloch bands. Multiplying the first equation by $\bar{A}$, the second by $\bar{B}$, and making use of the normalisation condition $|A|^2 + |B|^2 = 1$, yields the generalised eigenvalues
\begin{equation}
\label{eqn:chem pot}
\tilde\mu = cz^2 + i\gamma z - 2\cos k \left(\bar{A} B + \bar{B} A\right).
\end{equation}
For $k = \pm \frac{\pi}{2}$ equation (\ref{eqn:chem pot}) reduces to
\begin{equation}
\tilde\mu = cz^2 + i\gamma z
\end{equation}
and (\ref{eqn:nonlin mat}) provides the two conditions
\begin{equation}
\left(cz+i\gamma \right)A = \tilde\mu A, \quad -\left(cz+i\gamma \right) B = \tilde\mu B.
\end{equation}
In this case the only solutions are $z = \pm 1$, corresponding to $(A=1,B=0)$ and $(A=0,B=1)$ with generalised eigenvalues $\mu = g$ and $\mu = g - 2i\gamma$ respectively.

When $k \neq \pm \frac{\pi}{2}$ it can be shown that the possible values of $z$ are given by the real roots of the polynomial
\begin{equation}
\left(c^2+ \gamma^2\right)z^4 + \left(4\cos^2 k - c^2 - \gamma^2\right)z^2 = 0,
\end{equation}
which has the four solutions
\begin{equation}
\label{eqn:z sols}
z = 0,0,\pm \sqrt{1-\frac{4\cos^2 k}{c^2+ \gamma^2}}.
\end{equation}

Let us now parameterise the nonlinear Bloch states as
\begin{equation}
\label{eqn:A B params}
A = \sqrt{\frac{1+z}{2}}\ue^{-iv}, \quad B = \sqrt{\frac{1-z}{2}}\ue^{iv}.
\end{equation}
Following the analysis in \cite{Grae12b} it can be shown that when $k \neq \pm\frac{\pi}{2}$ there are up to four possible solutions with
\begin{numcases}{(z,v)=}\left(0,\frac{1}{2}\arcsin\left(\frac{\gamma}{2\cos k}\right)\right),\label{eqn:nl sols 1}\\
\left(\pm\sqrt{1-\frac{4\cos^2 k}{c^2+ \gamma^2}},-\frac{1}{2}\arctan\left(\frac{\gamma}{c}\right)\right),\label{eqn:nl sols 2}
\end{numcases}
where the phase coordinate for the second solution in (\ref{eqn:nl sols 1}) is given by $\pi/2 - v$. The nonlinear eigenvalues corresponding to the two solutions in (\ref{eqn:nl sols 1}) are given by
\begin{equation}
\label{eqn:nl eigs 1}
\mu_\mp = \mp \mathrm{sgn}(\cos k)\sqrt{4 \cos^2 k - \gamma^2} + c - i\gamma
\end{equation}
respectively. Those corresponding to the two solutions (\ref{eqn:nl sols 2}) may be written in the form
\begin{equation}
\label{eqn:nl eigs 2}
\mu = 2c\left[1 - \frac{2\cos^2 k}{c^2+\gamma^2}\left(1+\mathrm{sgn}(\cos k)\right)\right] + i\gamma(z-1).
\end{equation}
In the latter expression the real part of $\mu$ is constant when $\cos k$ is negative, resulting in nonlinear Bloch bands that are $2\pi$-periodic. This should be contrasted with the non-interacting ($g=0$) Bloch bands (Figure \ref{fig:spdisp}), which were found to be $\pi$-periodic.

 \begin{figure}[htp]
        \centering
              \includegraphics[width=0.49\textwidth]{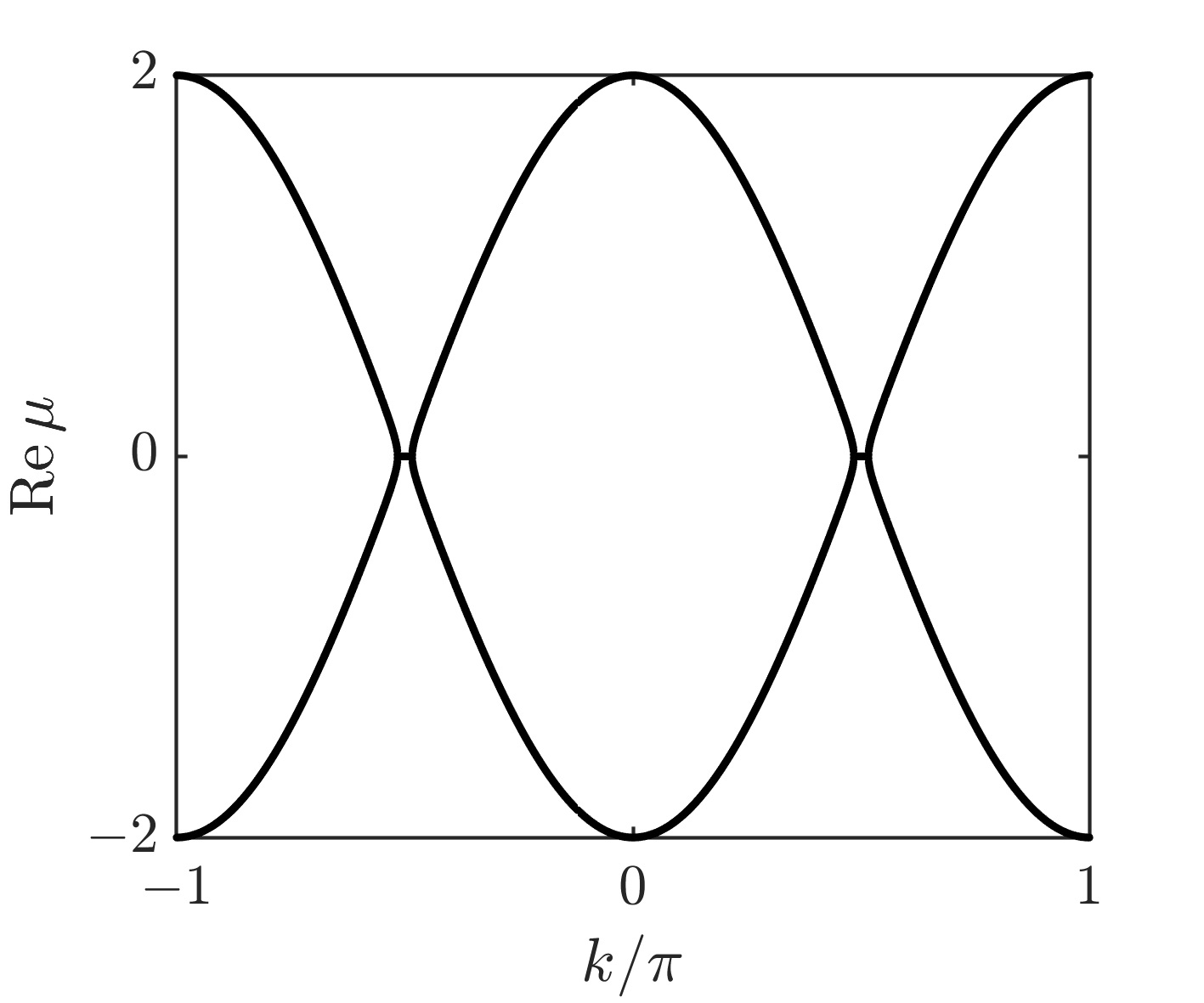} 
              \includegraphics[width=0.49\textwidth]{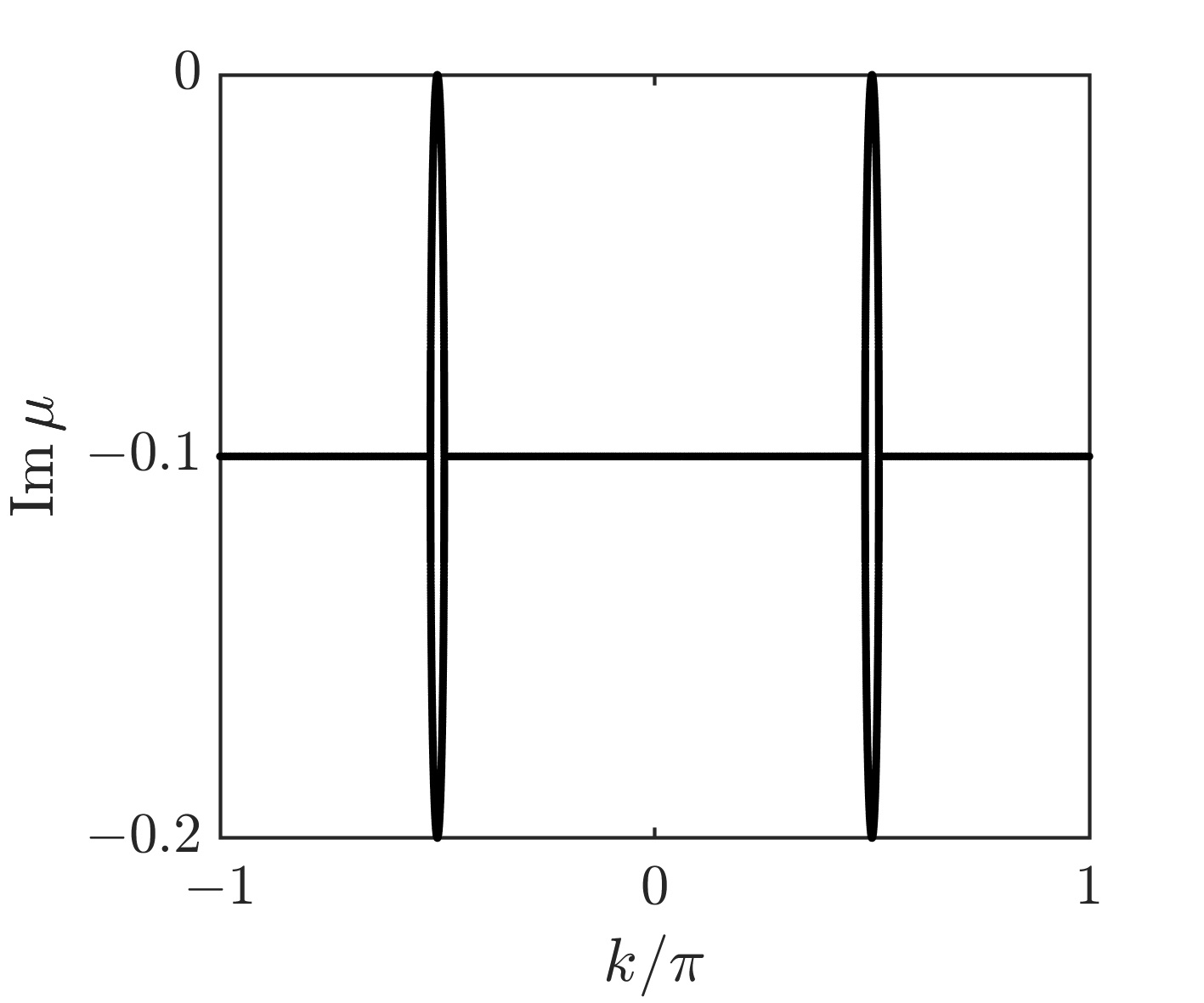} \hfill
              \includegraphics[width=0.49\textwidth]{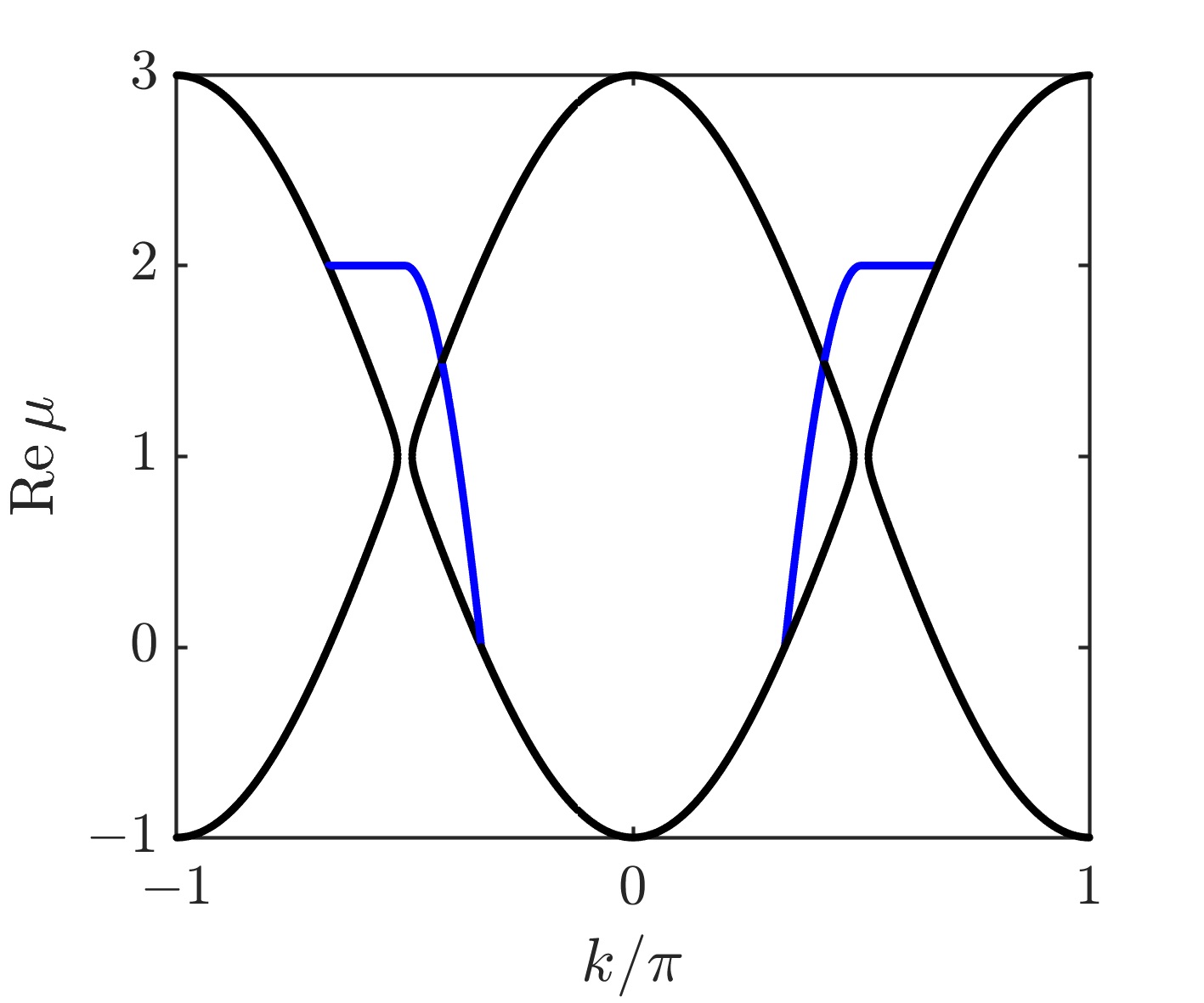} 
              \includegraphics[width=0.49\textwidth]{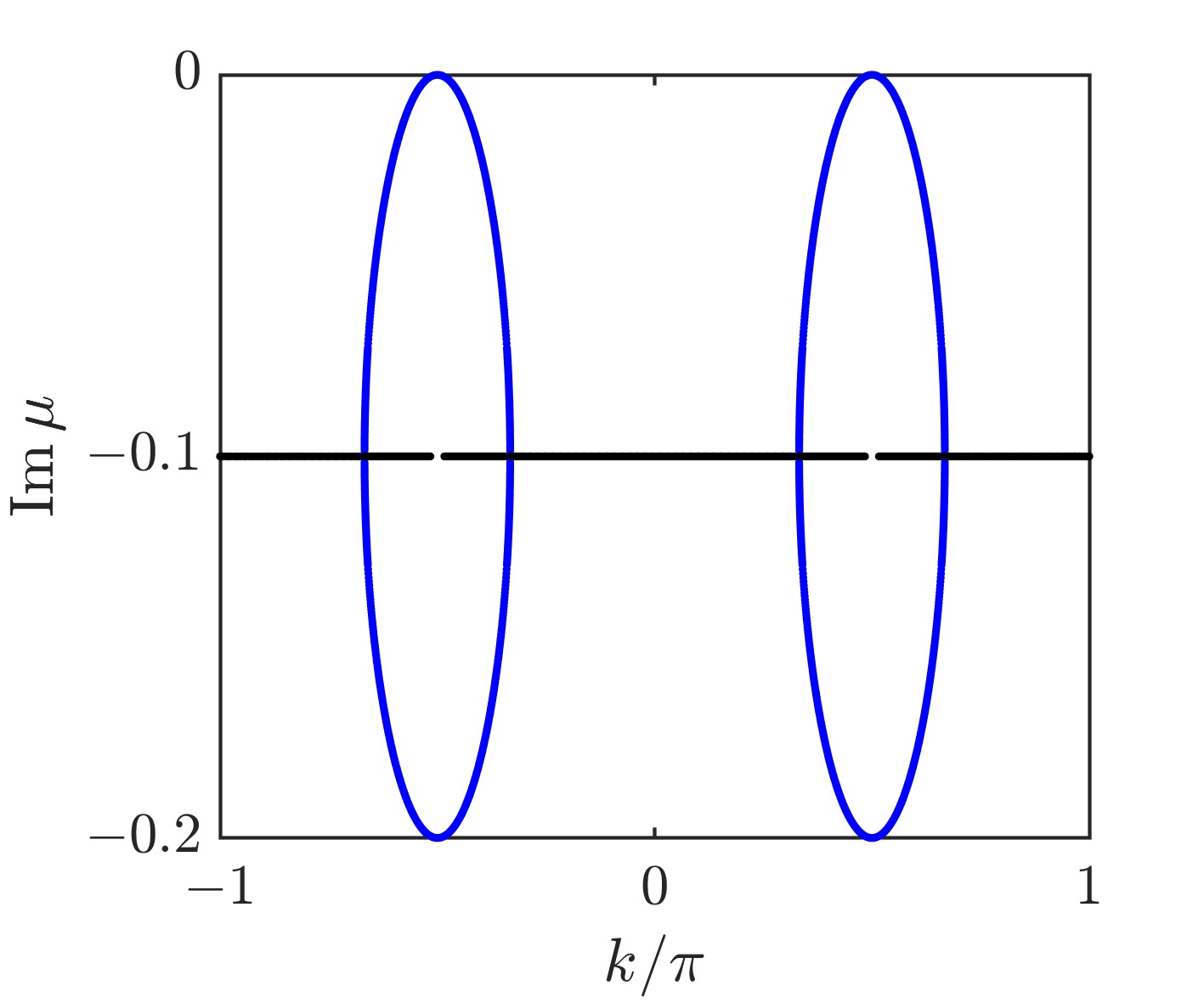} \hfill
              \includegraphics[width=0.49\textwidth]{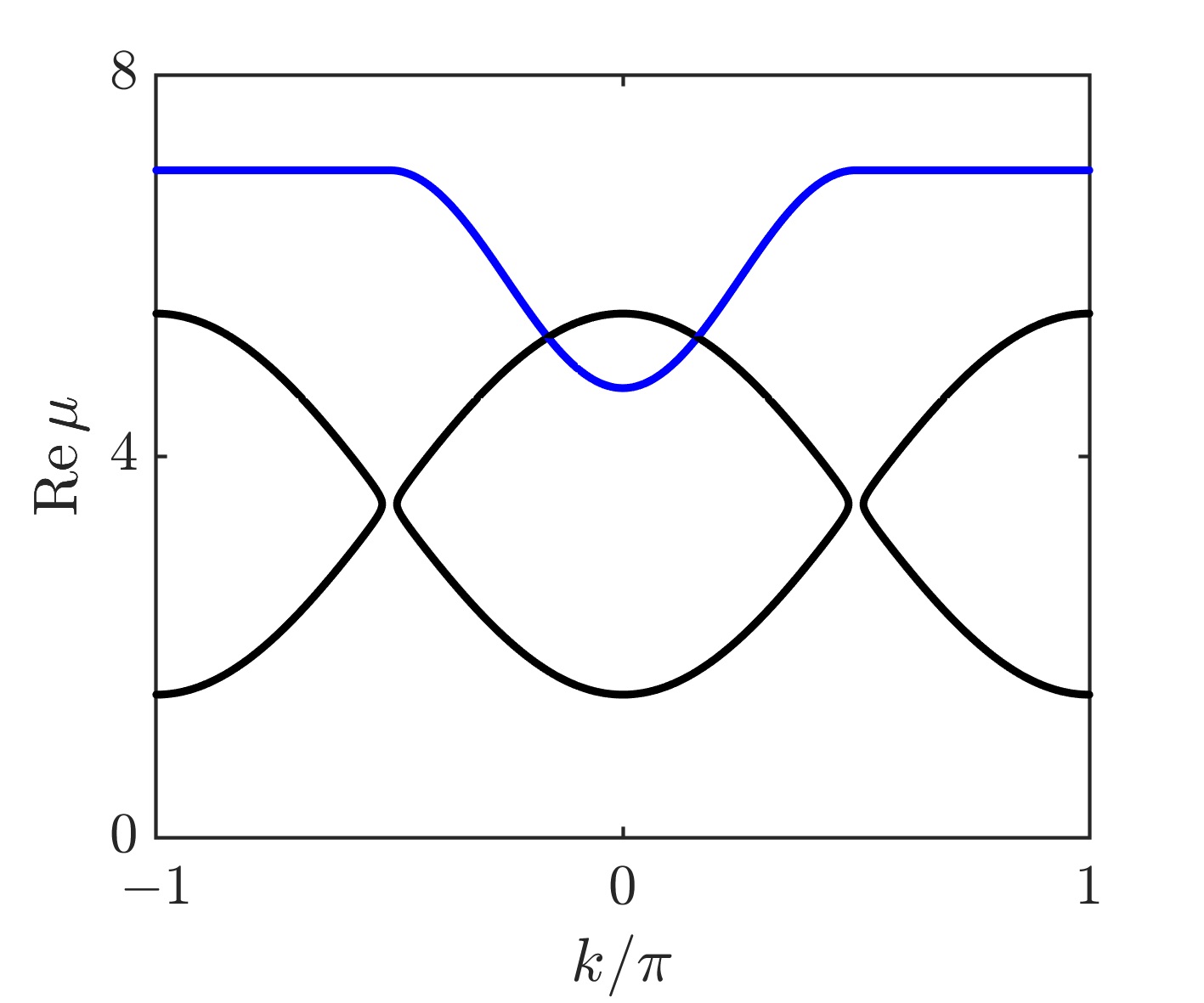} 
              \includegraphics[width=0.49\textwidth]{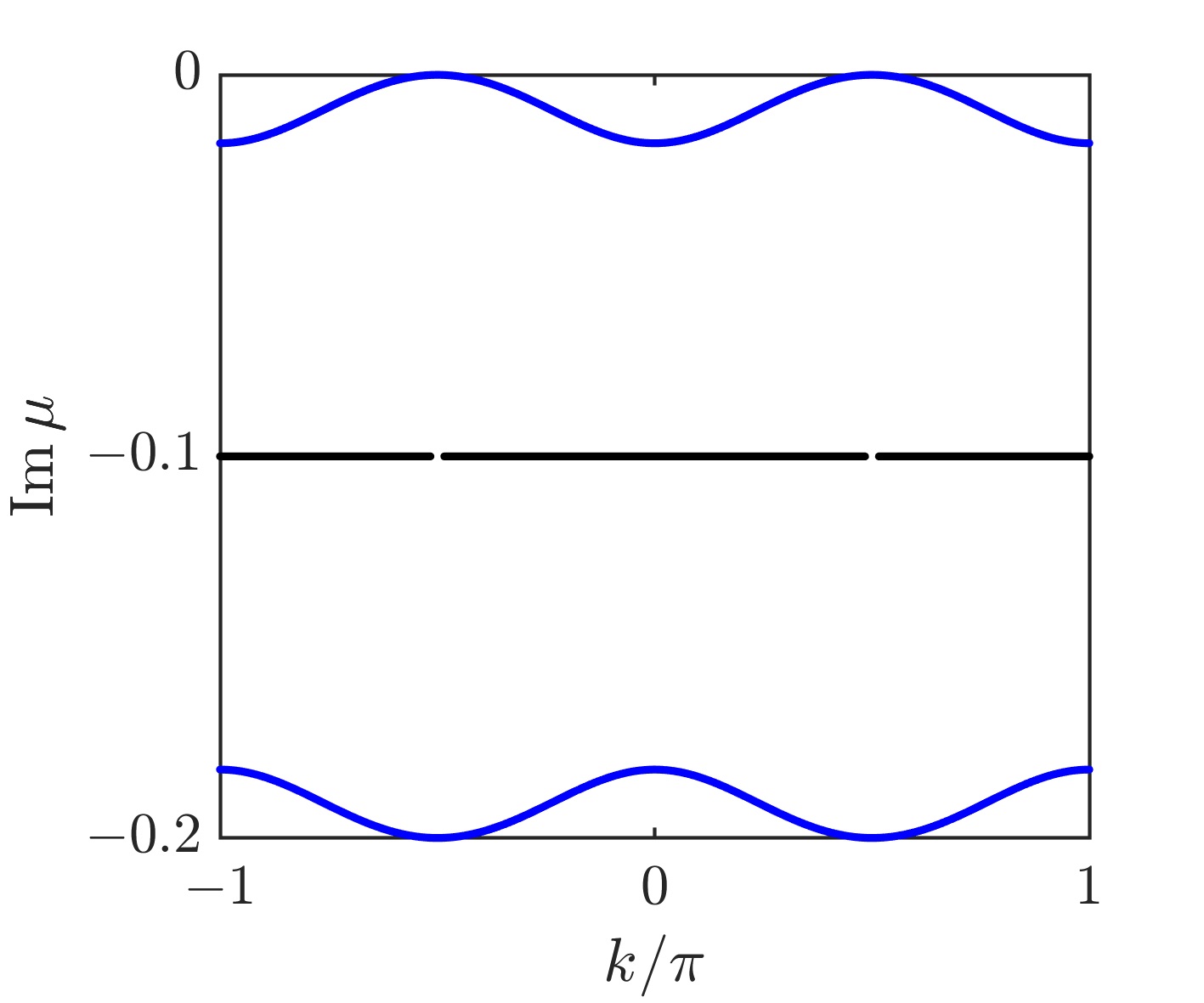} \hfill         
        \caption{Real (left column) and imaginary (right column) parts of the nonlinear Bloch bands for $g=0$ (top row), $g=2$ (middle row) and $g=7$ (bottom row) when $\gamma = 0.1$.} 
        \label{fig:nlbb}
\end{figure}

Only pairs $(z,v)$ that are real valued correspond to solutions. Thus, there are only two solutions for $|2\cos k|<\gamma$ and two for $4\cos^2 k > c^2 + \gamma^2$. There are three solutions at the exceptional points $2|\cos k| = \gamma$, where the two solutions (\ref{eqn:nl sols 1}) coalesce in an exceptional point (EP2). The nonlinearity introduces additional exceptional points at the quasimomentum values $2|\cos k| = \sqrt{c^2+\gamma^2}$. At these points there can be three or two solutions. There are three solutions for $\mathrm{sgn}\left(\cos k\right) > 0$, as the two solutions in (\ref{eqn:nl sols 2}) coalesce (EP2). Surprisingly, there are only two solutions for $\mathrm{sgn}\left(\cos k\right) <0$, as both the solutions in (\ref{eqn:nl sols 2}) coalesce with one solution in (\ref{eqn:nl sols 1}) (EP3). Otherwise there are four solutions. The existence of an EP2, EP3 and even an EP4 has previously been demonstrated for a BEC in a PT-symmetric double well potential \cite{Dizd15,Dast13,Heis13}.

The real and imaginary parts of $\mu$ are plotted in Figure \ref{fig:nlbb} for weak ($g = 2$) and strong ($ g = 7$) interaction strengths, when the decay rate is $\gamma = 0.1$. The Bloch bands for the non-interacting ($g=0$) case are also shown in the top row for comparison. The combination of interaction and dissipation leads to significant changes in the band structure compared to the single particle (non-interacting) case. When $g$ is switched on a pair of solutions vanish just after the exceptional points at $2|\cos k| = \gamma$, visible around $k = \pm \pi/2$, and a new band emerges (coloured blue). The interaction introduces additional exceptional points at $2|\cos k| = \sqrt{c^2+\gamma^2}$ that vanish when $\sqrt{c^2+\gamma^2}>2$. Away from exceptional points the solutions in the blue band have the same real part but different imaginary parts. The solutions in the black band have different real parts but degenerate imaginary parts. 

\subsection{Stability analysis}
In contrast to the Bloch states of the single-particle system, the nonlinear Bloch states can become dynamically unstable. The stability of the stationary states can be determined by examining the effect of a small time-dependent perturbation of the stationary solution of the form \cite{Witt11}
\begin{equation}
\label{eqn:bdg pert}
\psi_j(t) = \ue^{-i\mu t}\phi_j + \left\{
                \begin{array}{ll}
                 \ue^{ikj-i\mu t}\left(\ue^{- i\omega t}\delta \chi^A_- + \ue^{i\bar{\omega}t}\overbar{\delta \chi}^A_+\right), \quad j \,\, \textnormal{even}, \\
                  \ue^{ikj-i\mu t}\left(\ue^{- i\omega t}\delta \chi^B_- + \ue^{i\bar{\omega}t}\overbar{\delta \chi}^B_+\right), \quad j \,\, \textnormal{odd}.
                \end{array}
              \right.
\end{equation}
By inserting the ansatz (\ref{eqn:bdg pert}) into the field-free version of (\ref{eqn:cdnls tilt}) the problem reduces to the two-mode system analysed in \cite{Grae12b}, and the stability analysis obtained therein can immediately be applied. In particular, it is sufficient to consider the nonlinear Schr\"odinger equation 
\begin{equation}
\label{eqn:nl schrodinger}
i\frac{d}{dt}\begin{pmatrix}\Psi_1 \\ \Psi_2 \end{pmatrix} = H_{nl}(|\Psi_1|^2,|\Psi_2|^2)\begin{pmatrix}\Psi_1\\ \Psi_2 \end{pmatrix},
\end{equation}
where $H_{nl}(|\Psi_1|^2,|\Psi_2|^2)$ is the two-level nonlinear Hamiltonian
\begin{equation}
\label{eqn:nonlinear Ham}
H_{nl}(|\Psi_1|^2,|\Psi_2|^2) = \begin{pmatrix} g |\Psi_1|^2 & -2\cos k\\-2\cos k & g |\Psi_2|^2 - 2i\gamma \end{pmatrix}.
\end{equation}
In the previous section we found that the stationary states of this system are given by $\chi = (A,B)$, with corresponding nonlinear eigenvalues $\mu$. Therefore, the stability analysis (\ref{eqn:bdg pert}) reduces to analysing the linearised effect of a small time-dependent perturbation to the stationary solution $\chi$ of the form
\begin{equation}
\Psi(t) = \ue^{-i\mu t}\left(\chi + \delta \chi_- \ue^{-i\omega t} + \overbar{\delta \chi}_+ \ue^{i\bar{\omega} t}\right),
\end{equation}
where the components of $\delta \chi_\pm = (\delta \chi^A_\pm,\delta \chi^B_\pm)$ are defined in (\ref{eqn:bdg pert}). In this case it can be shown \cite{Grae12b} that the Bogoliubov-de Gennes equations yield the eigenvalue equation
\begin{equation}
\label{eqn:bdg eigenval prob}
M\begin{pmatrix}\delta \chi_-\\ \delta \chi_+ \end{pmatrix} = \omega \begin{pmatrix} \delta \chi_- \\ \delta \chi_+ \end{pmatrix},
\end{equation}
where $M$ is the $4 \times 4$ matrix
\begin{equation}
\label{eqn:bdg M mat}
M = \begin{pmatrix} H^0_{nl} - \mu + cP\begin{pmatrix}|A|^2 & -A\bar{B}\\-\bar{A}B & |B|^2  \end{pmatrix}P & cP \begin{pmatrix} A^2 & -AB \\ -AB & B^2 \end{pmatrix}\bar{P} \\ -c\bar{P}\begin{pmatrix} \bar{A}^2 &- \bar{A}\bar{B}\\-\bar{A}\bar{B} & \bar{B}^2\end{pmatrix}P & -\bar{H}^0_{nl} + \bar{\mu} - c\bar{P}\begin{pmatrix}|A|^2 & -\bar{A}B\\ -A\bar{B} & |B|^2\end{pmatrix}\bar{P}\end{pmatrix}.
\end{equation}
Here $H^0_{nl}$ is the nonlinear Hamiltonian (\ref{eqn:nonlinear Ham}) evaluated at the stationary solution $\chi = (A,B)$ and $P$ is the projection matrix orthogonal to $\chi$,
\begin{equation}
P = \begin{pmatrix} 1-|A|^2 & -A\bar{B}\\-\bar{A}B & 1-|B|^2 \end{pmatrix}.
\end{equation}

 \begin{figure}[h]
 
        \centering
             \includegraphics[width=0.49\textwidth]{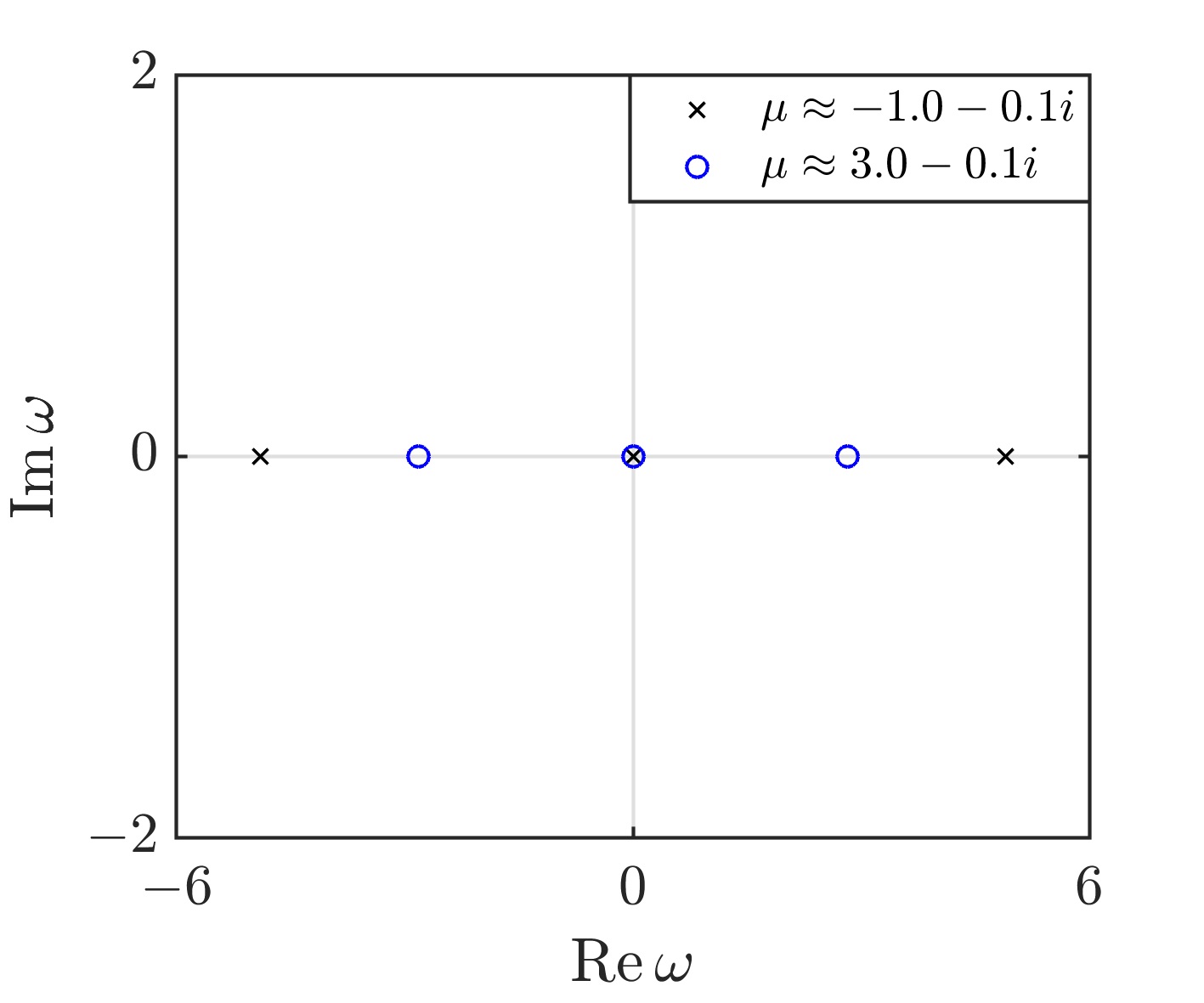}
             \includegraphics[width=0.49\textwidth]{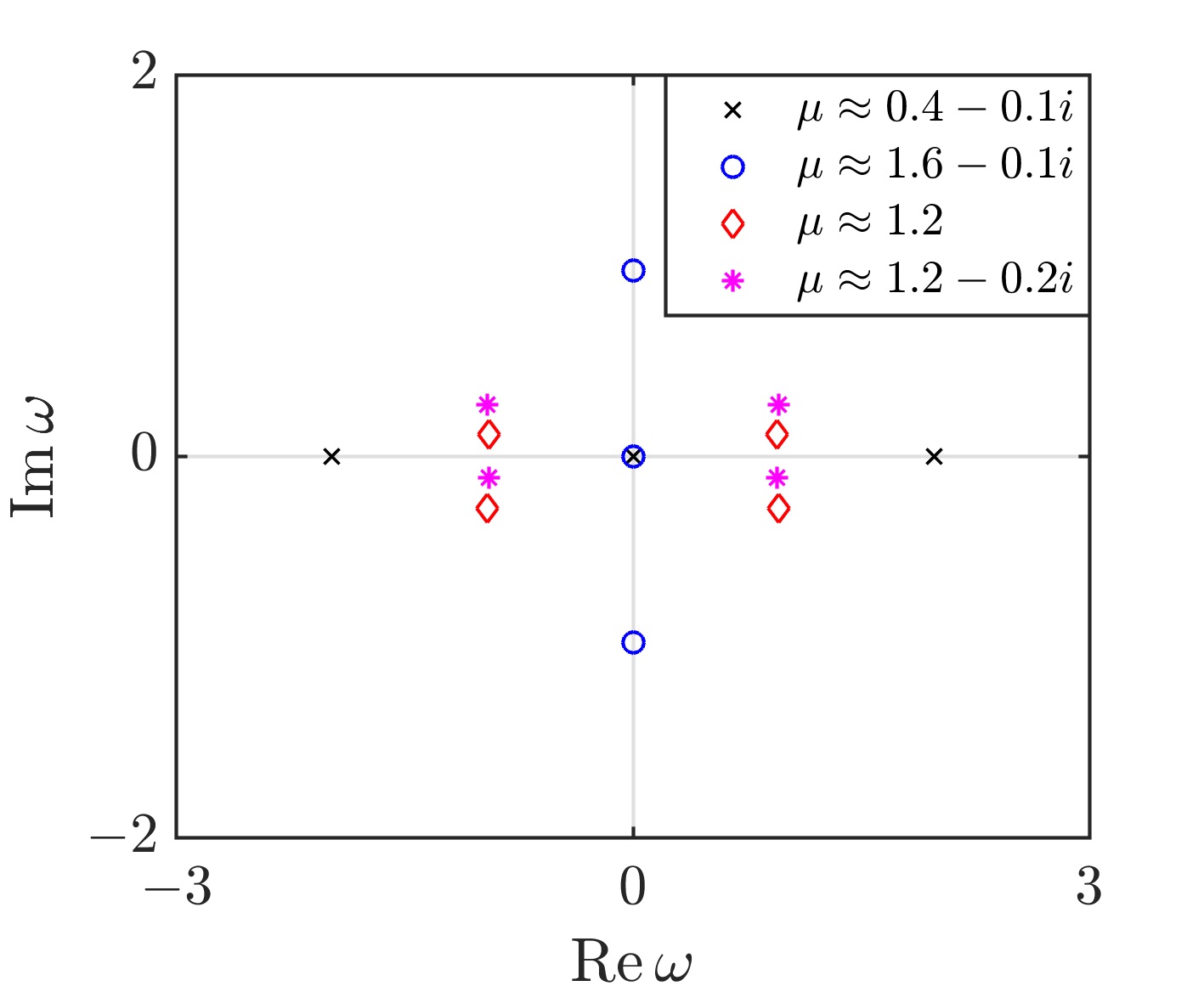}\\
              \includegraphics[width=0.49\textwidth]{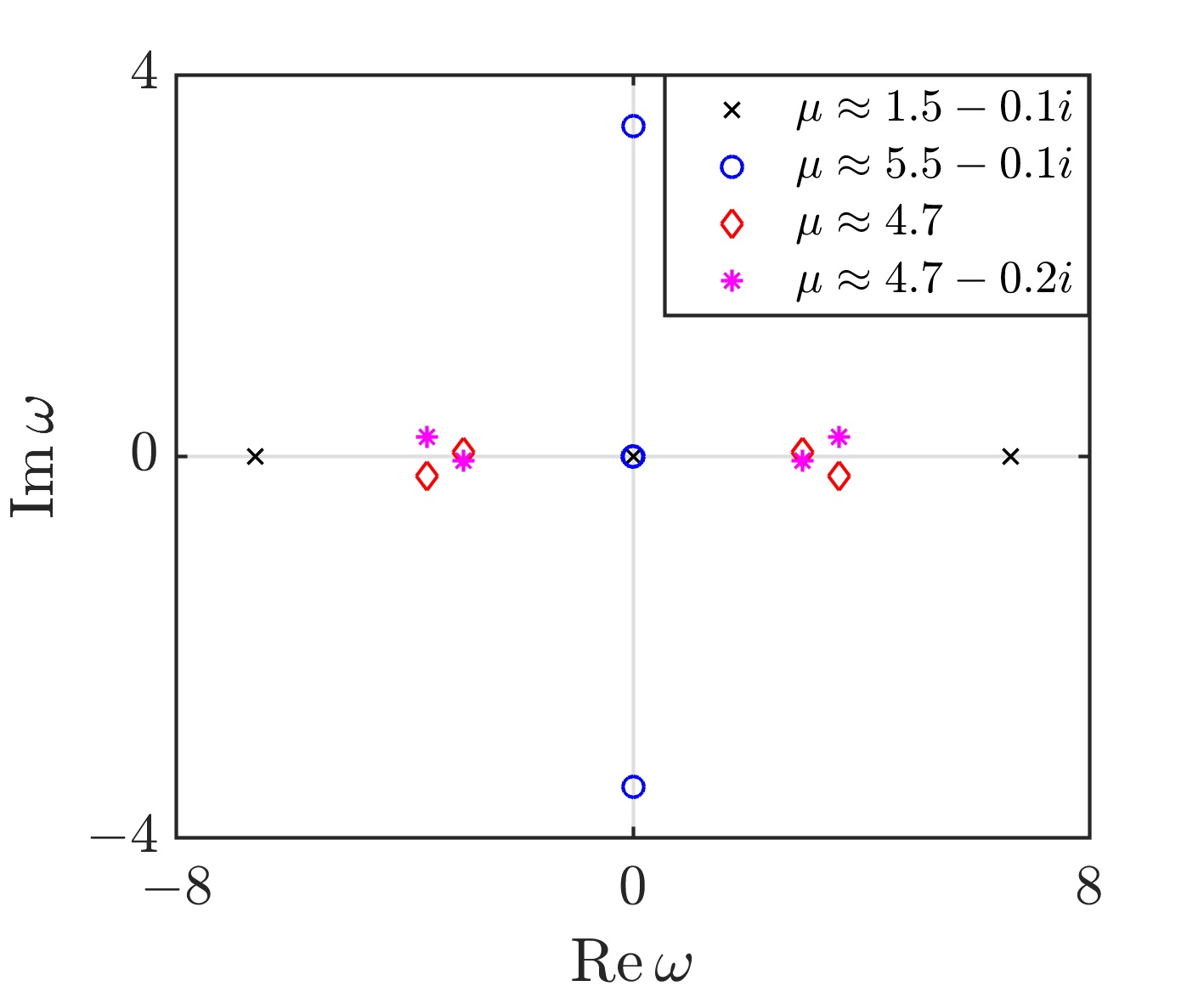}
             \includegraphics[width=0.49\textwidth]{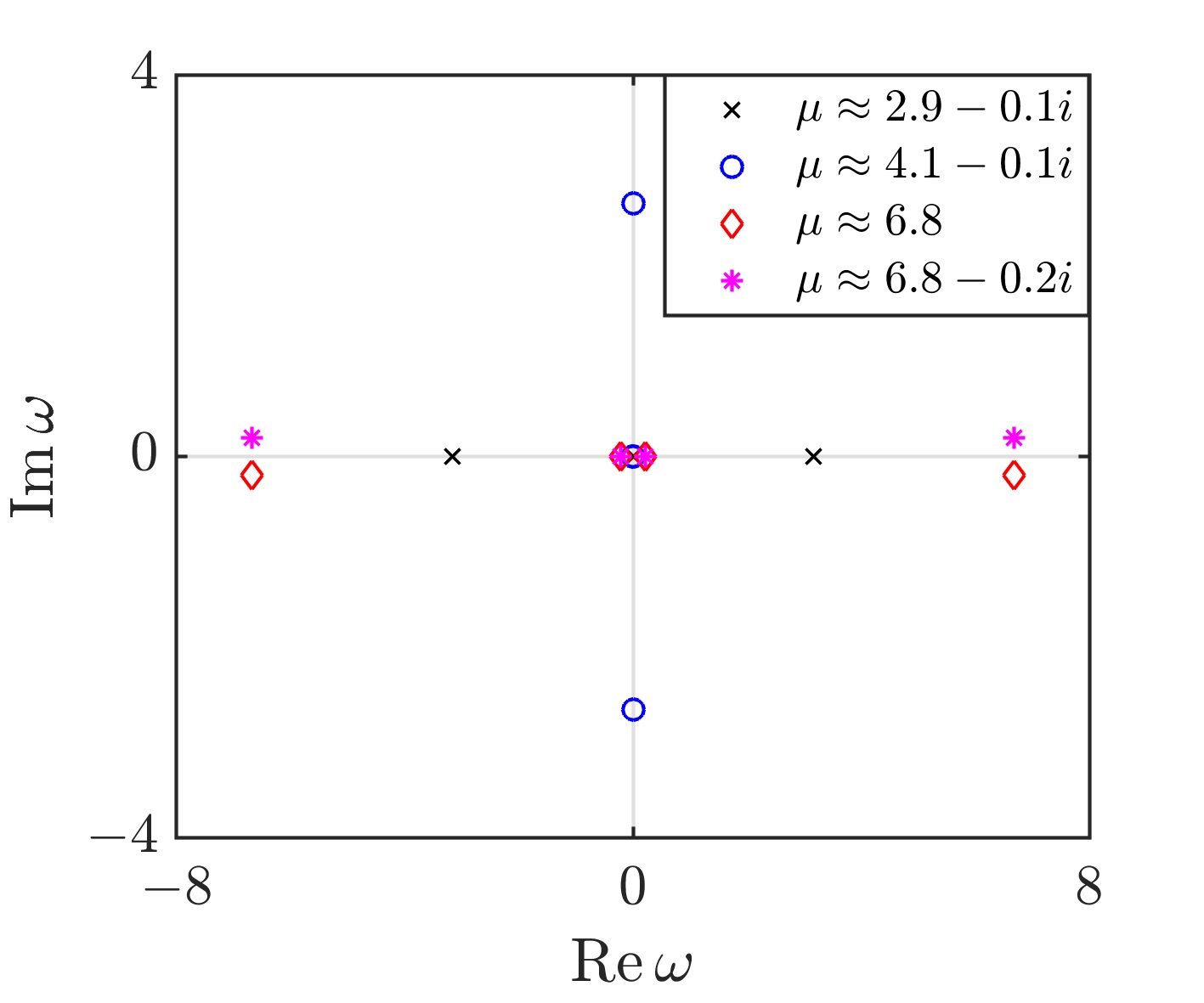}
             \caption{Eigenvalues $\omega$ of the stability matrix $M$ corresponding to different stationary states for $\gamma = 0.1$, $g=2$ (top row) and $g = 7$ (bottom row). The columns show results for different values of the quasimomentum, with $k=0$ (left) and $k=-0.4\pi$ (right).} 
        \label{fig:stability fixed}
\end{figure}

\begin{figure}[htb]
              \includegraphics[width=0.49\textwidth]{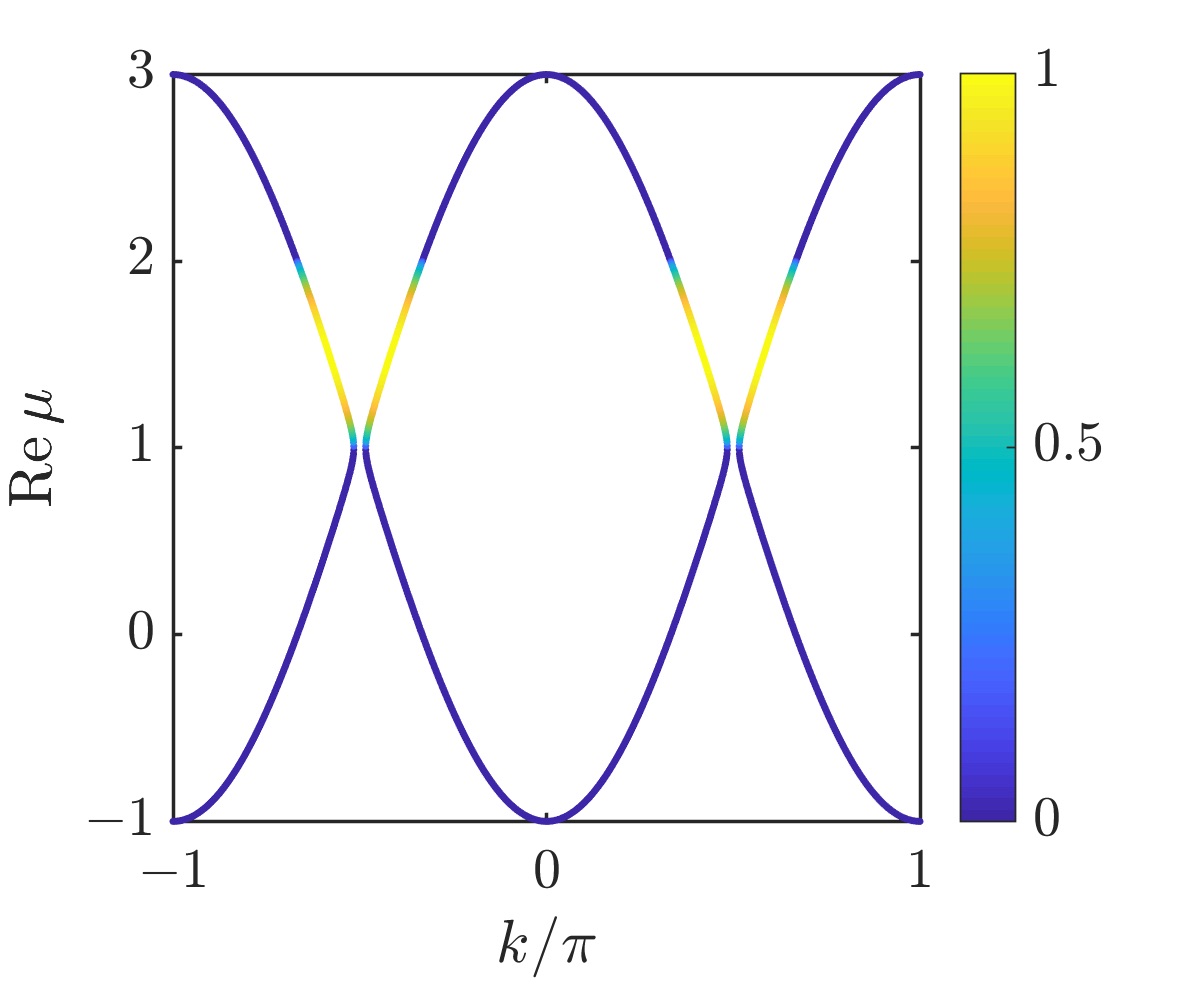} 
              \includegraphics[width=0.49\textwidth]{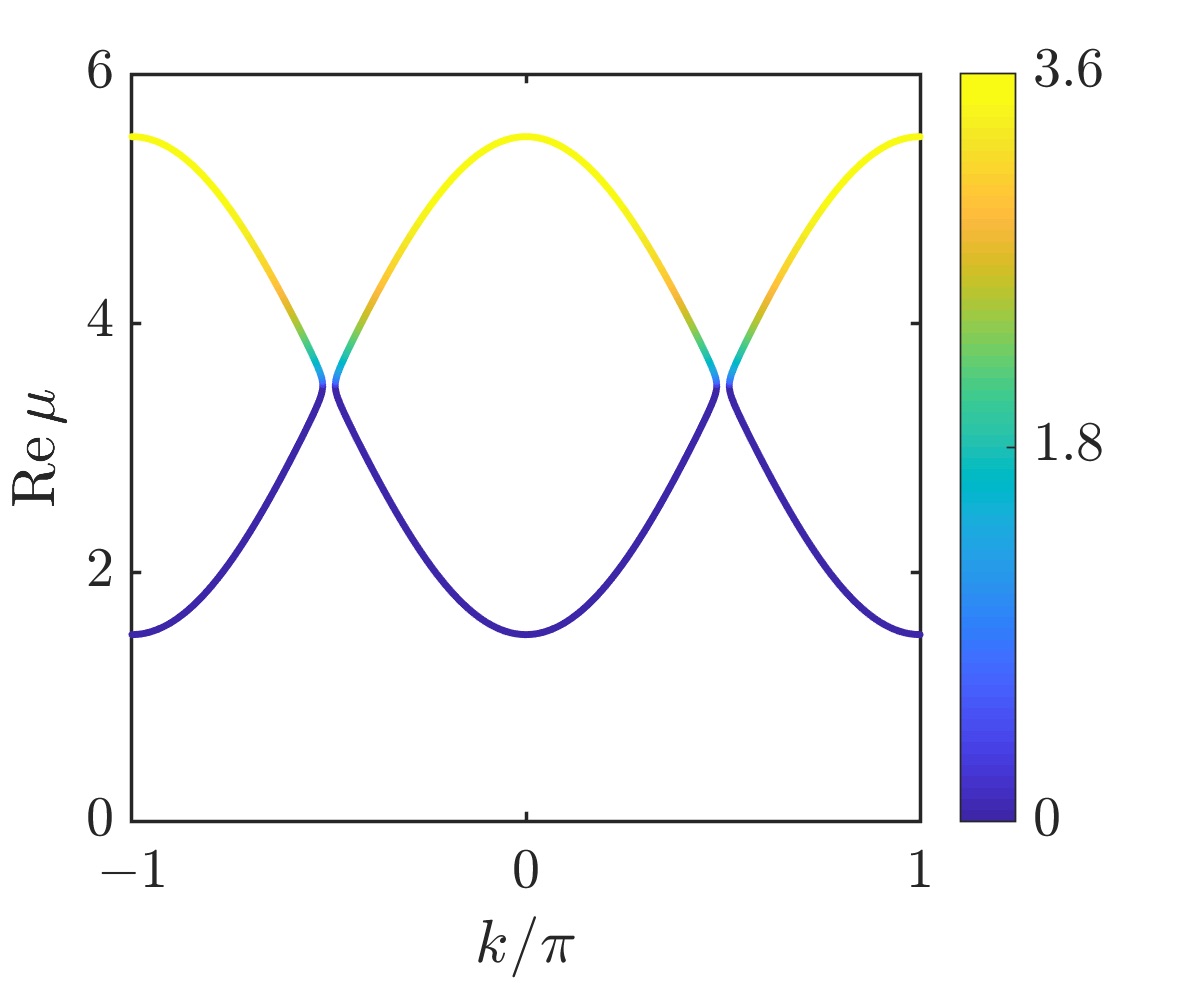}
              \includegraphics[width=0.49\textwidth]{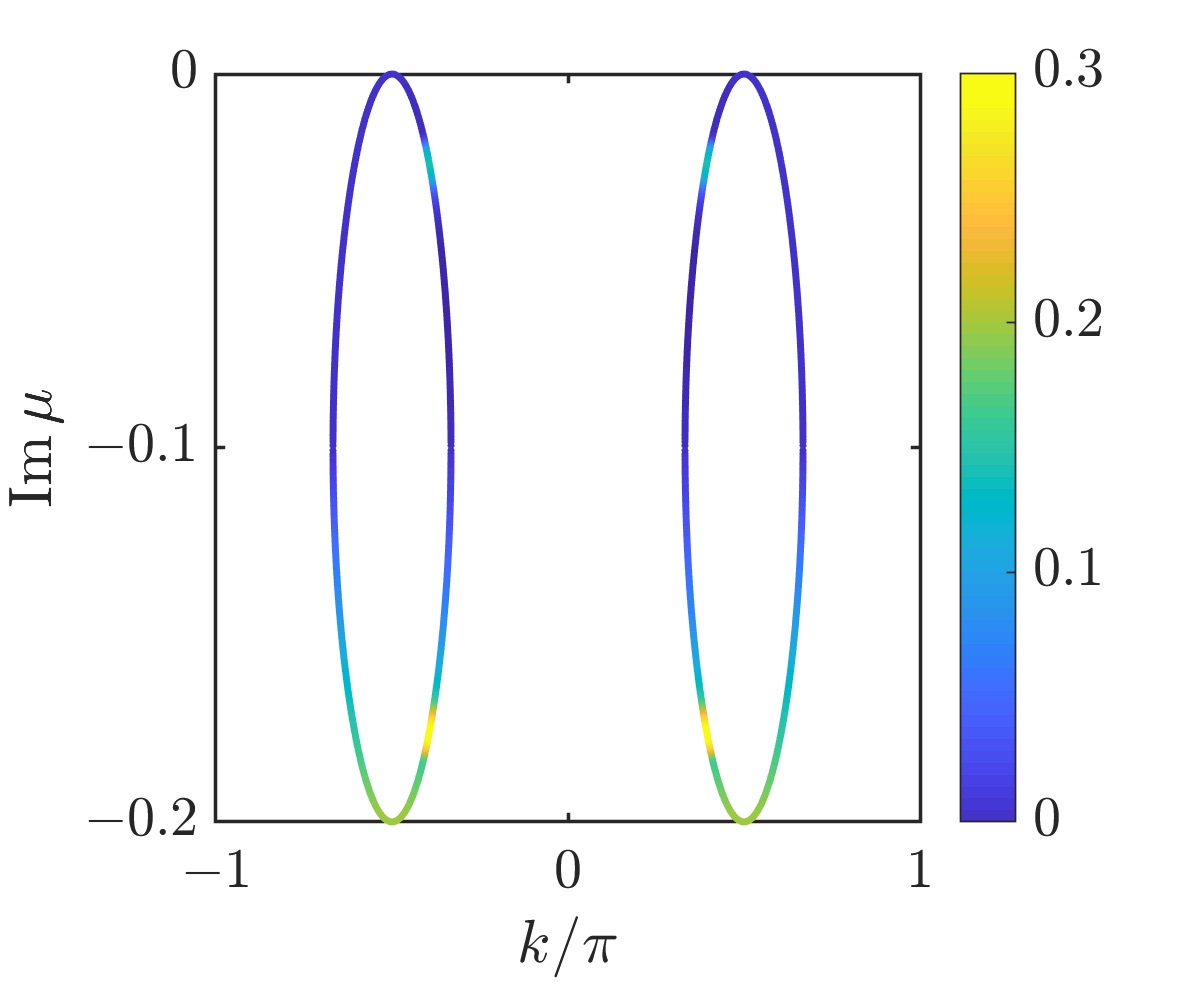} 
              \includegraphics[width=0.49\textwidth]{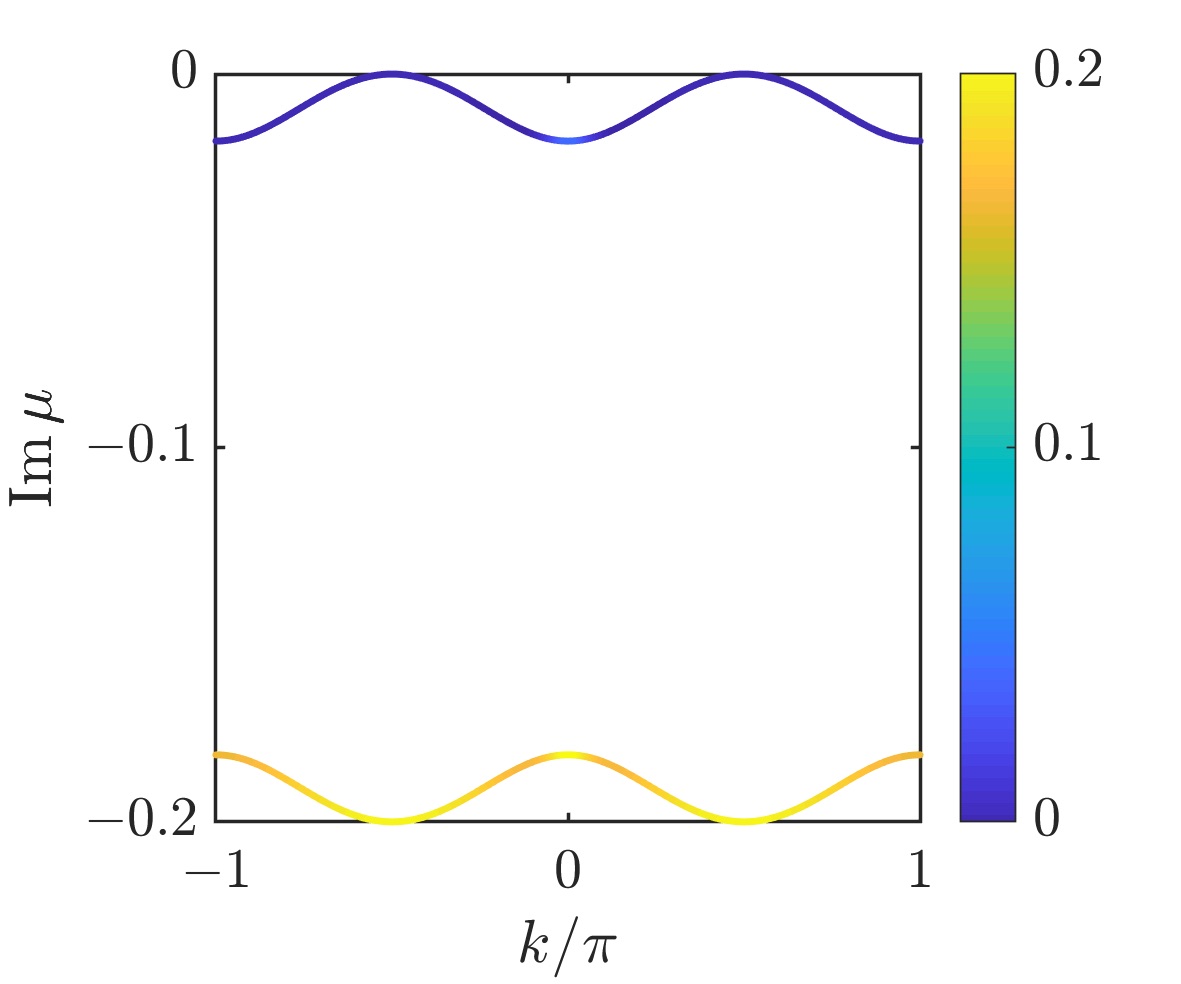}
             \caption{(Top row) Real parts of the nonlinear Bloch bands corresponding to the solutions (\ref{eqn:nl sols 1}). (Bottom row) Imaginary parts of the nonlinear Bloch bands corresponding to the solutions (\ref{eqn:nl sols 2}). False colours show $\max [\Im \omega]$, thus illustrating the stability of the solutions. In each row $\gamma = 0.1$, $g = 2$ (left column) and $g=7$ (right column).} 
        \label{fig:stability eigs}
\end{figure}

The imaginary parts of the eigenvalues $\omega$ determine the stability of the stationary solution $\chi$. When $\Im \omega \leq 0$ the perturbations are decaying or oscillating, and the solution is stable. On the other hand, if $\Im \omega > 0$ the perturbations grow exponentially and the solution is unstable. A short calculation reveals that at $k = \pm \pi/2$ the solution $\chi = (1,0)$ with $\mu = g$ is stable, while the second solution $\chi = (0,1)$ with $\mu = g - 2i\gamma$ is unstable. In general the stability of the stationary solutions changes as the system parameters are varied. This is illustrated in Figure \ref{fig:stability fixed} and Figure \ref{fig:stability eigs} for two different values of $g$ and varying quasimomentum values, with $\gamma = 0.1$. 

For the solutions (\ref{eqn:nl sols 1}), with corresponding nonlinear eigenvalues (\ref{eqn:nl eigs 1}), the eigenvalue equation (\ref{eqn:bdg eigenval prob}) yields the characteristic polynomial
\begin{equation}
\omega^4 + 4\left(\gamma^2 - 4\cos^2 k \mp \mathrm{sgn}(\cos k) c\sqrt{4\cos^2 k - \gamma^2}\right)\omega^2 = 0.
\end{equation}
It follows that for $\mathrm{sgn}(\cos k) > 0$ the solution corresponding to $\mu_+$ becomes unstable for quasimomentum values satisfying $4\cos^2 k < c^2 + \gamma^2$, while the solution corresponding to $\mu_-$ remains stable. The behaviour of these solutions is reversed when $\mathrm{sgn}(\cos k) < 0$, as illustrated in the top row of Figure \ref{fig:stability eigs}. The stability of these solutions is completely determined by the interaction strength $g$. On the other hand, changing $g$ and/or $\gamma$ can change the stability of the solutions (\ref{eqn:nl sols 2}). The change in stability as $g$ is increased is shown in the bottom row of Figure \ref{fig:stability eigs}. 

It is important to note that the stability obtained from the Bogoliubov-de Gennes equation can change during the time evolution of the system. Due to particle losses the norm ($|A|^2+|B|^2$) is no longer conserved. This effectively leads to a time-dependent interaction strength and, therefore, different stability behaviour during the time evolution. We shall now demonstrate that the stability of the nonlinear Bloch bands can have a strong influence on the mean-field dynamics, and even appears to influence the particle number expectation value dynamics of a two-particle quantum system.

\subsection{Mean-field and two-particle dynamics}
We now examine the dynamics of an initial pure BEC state (\ref{eqn:bec state}). The parameters $\psi_j$ of the BEC state are taken to be the components of a nonlinear stationary state $\phi_j$ (\ref{eqn:nl A and B}), weighted by a Gaussian envelope $\psi_j \propto \phi_j \ue^{-(j-x_0)^2/2\sigma^2}$. In particular, we assume that $k_0=0$ and consider the two stationary solutions in (\ref{eqn:nl sols 1}).

 \begin{figure}[htb]
        \centering
              \includegraphics[width=0.328\textwidth]{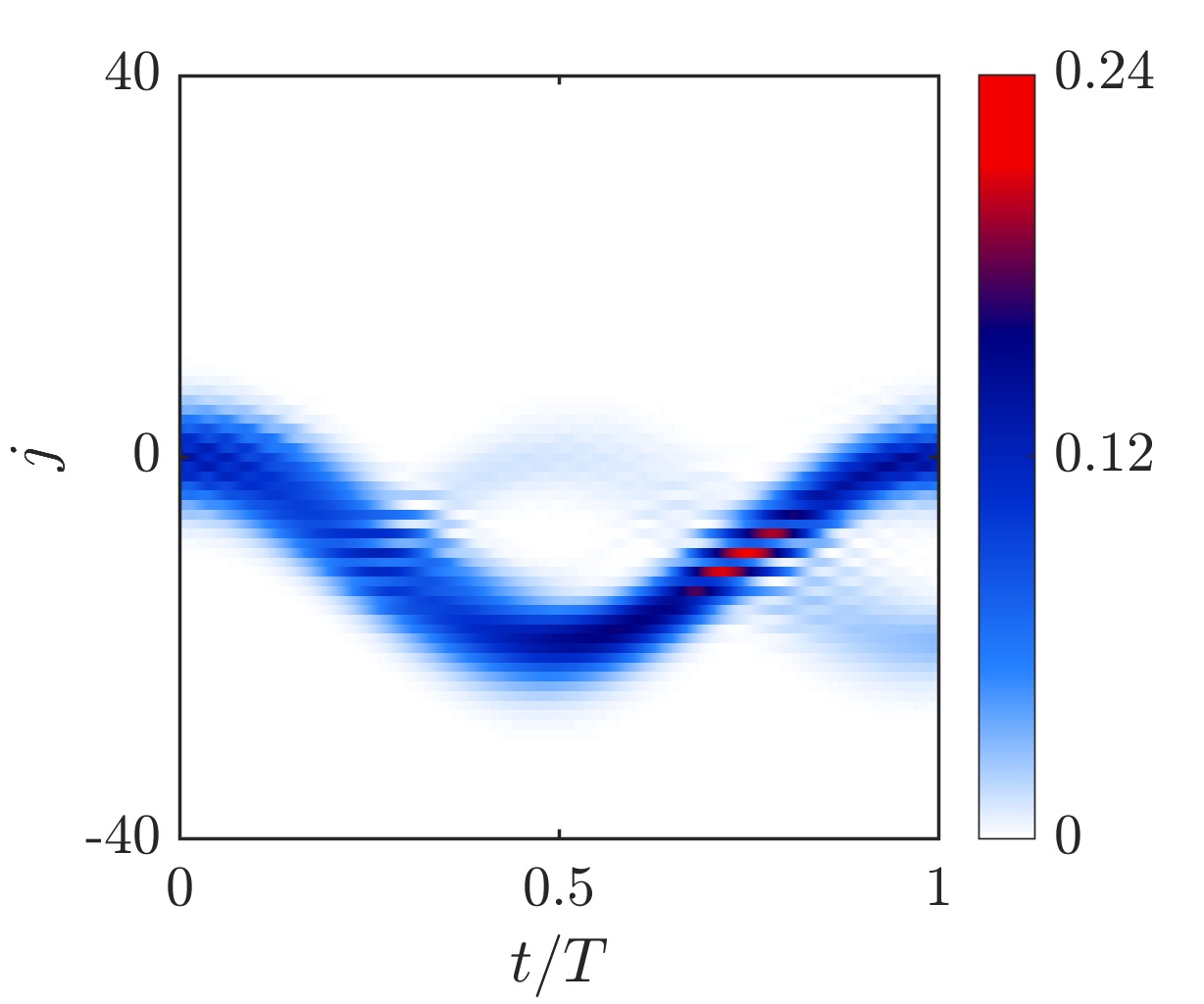} 
              \includegraphics[width=0.328\textwidth]{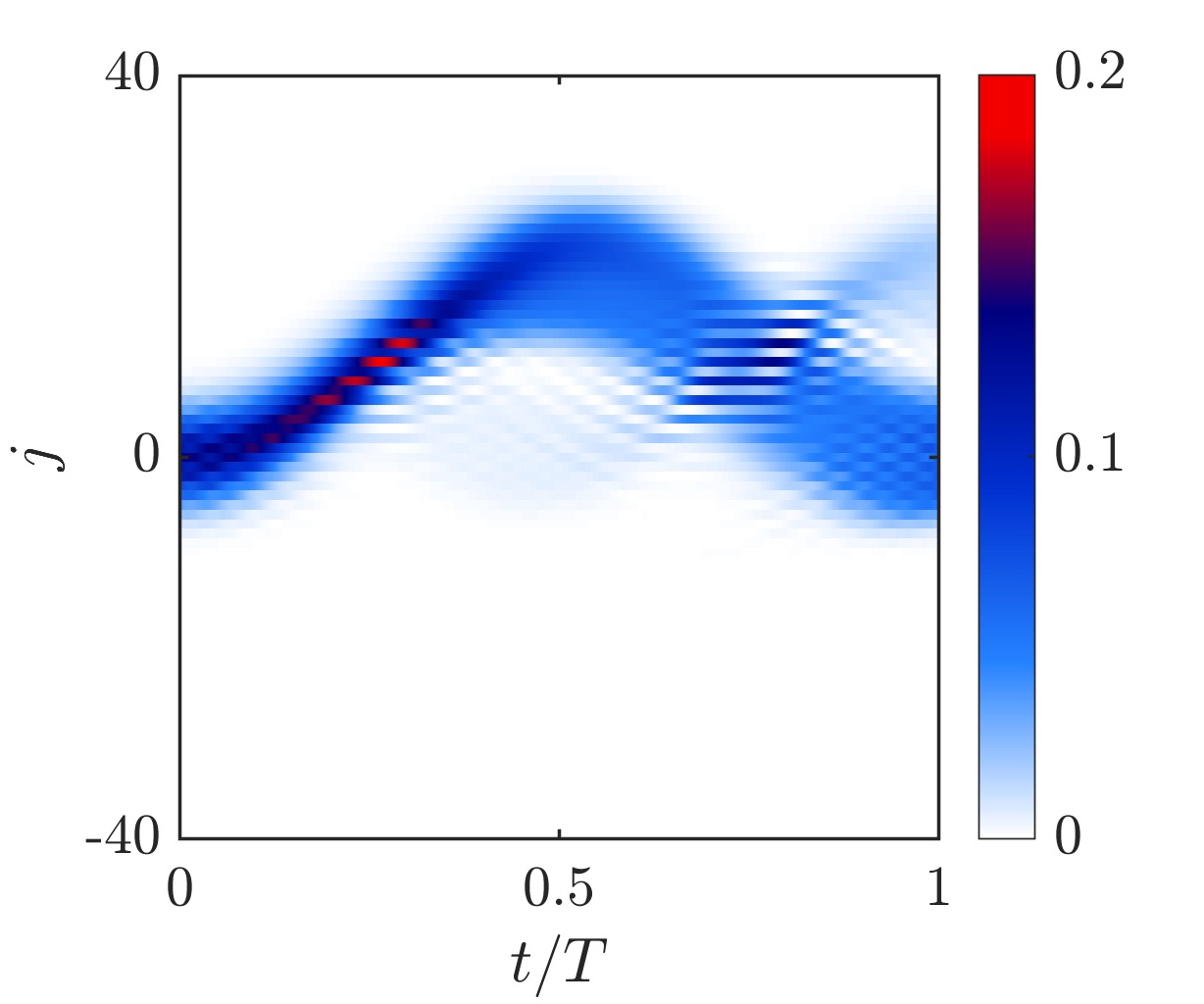} 
              \includegraphics[width=0.328\textwidth]{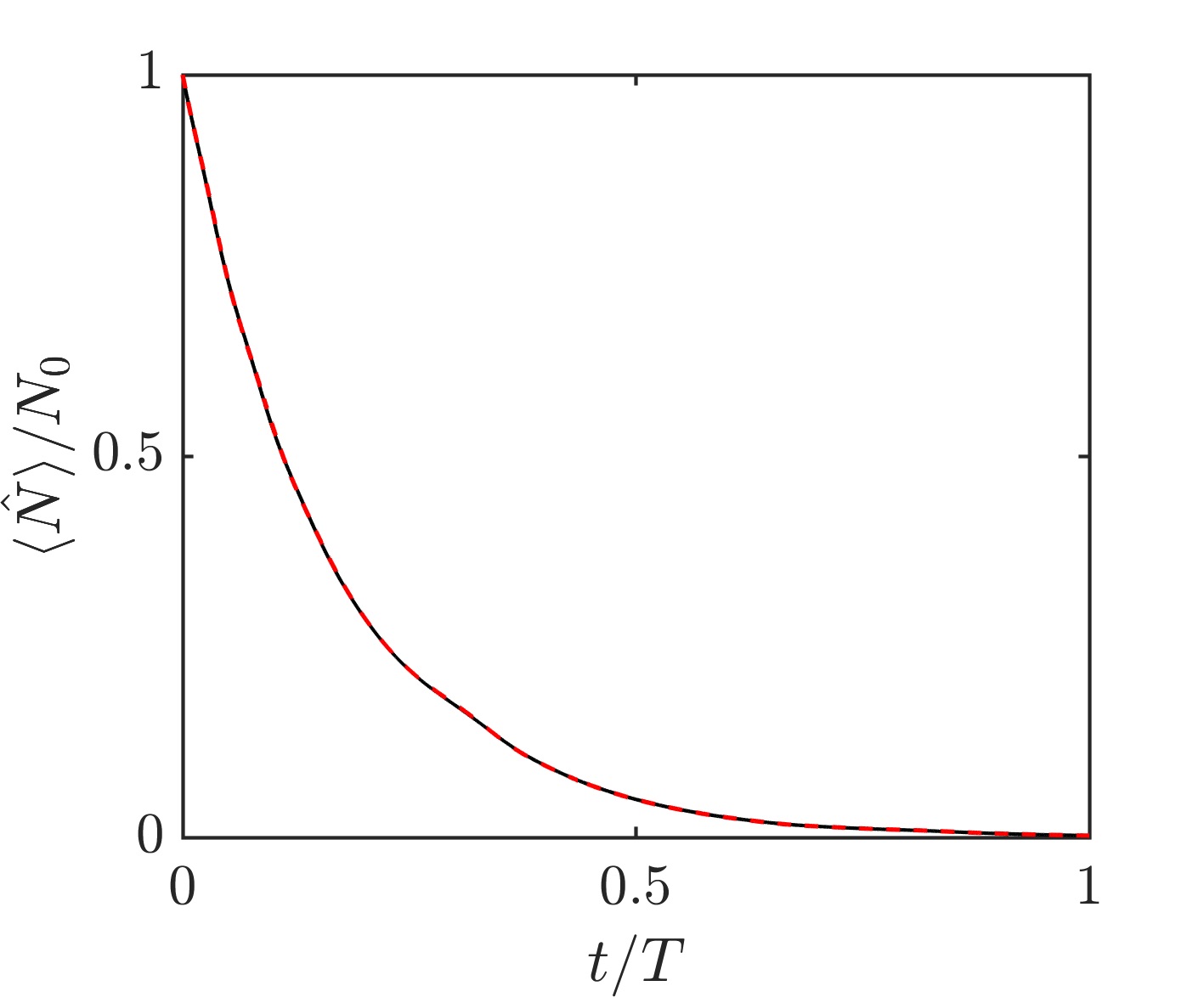}\\
              \includegraphics[width=0.328\textwidth]{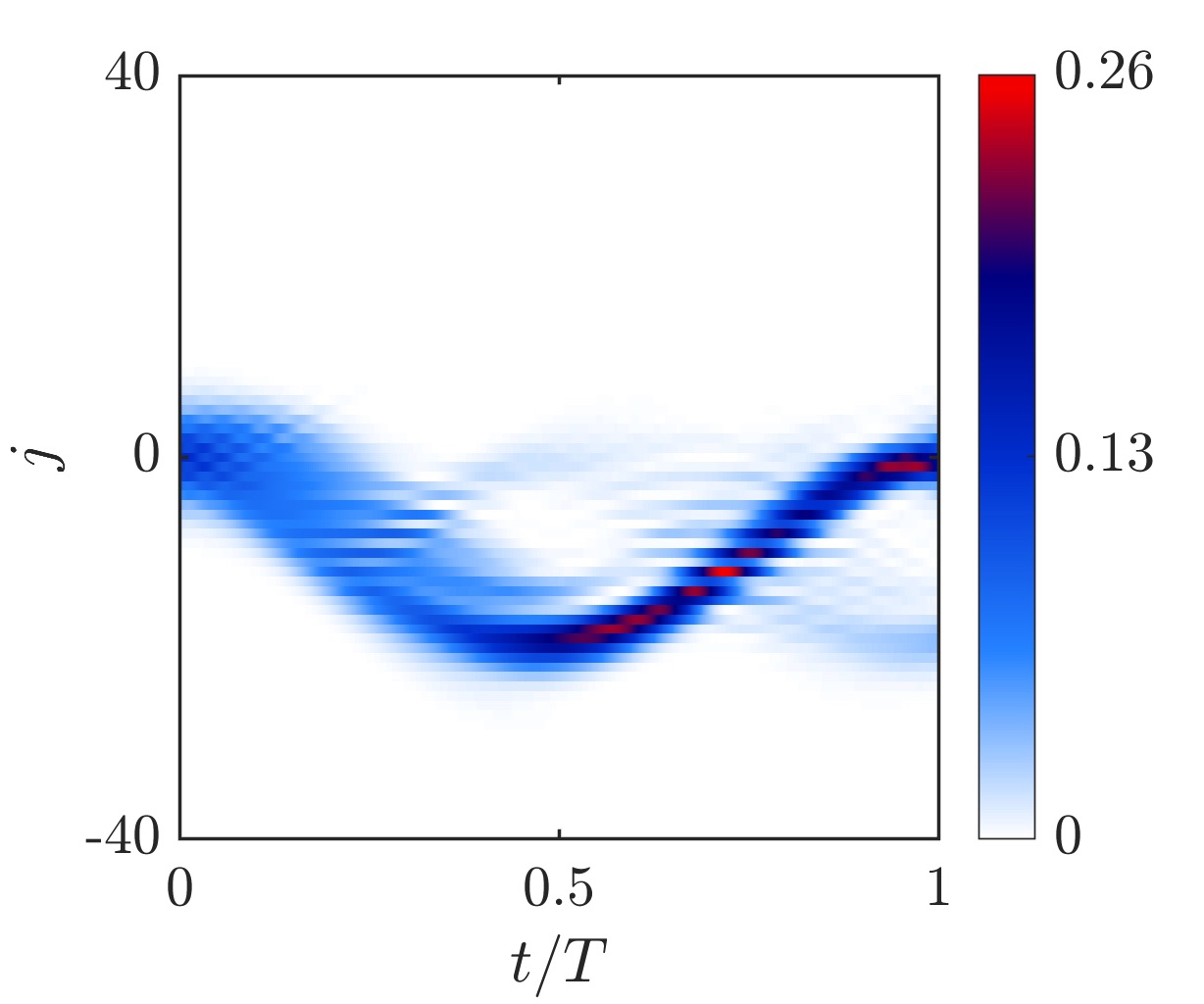} 
              \includegraphics[width=0.328\textwidth]{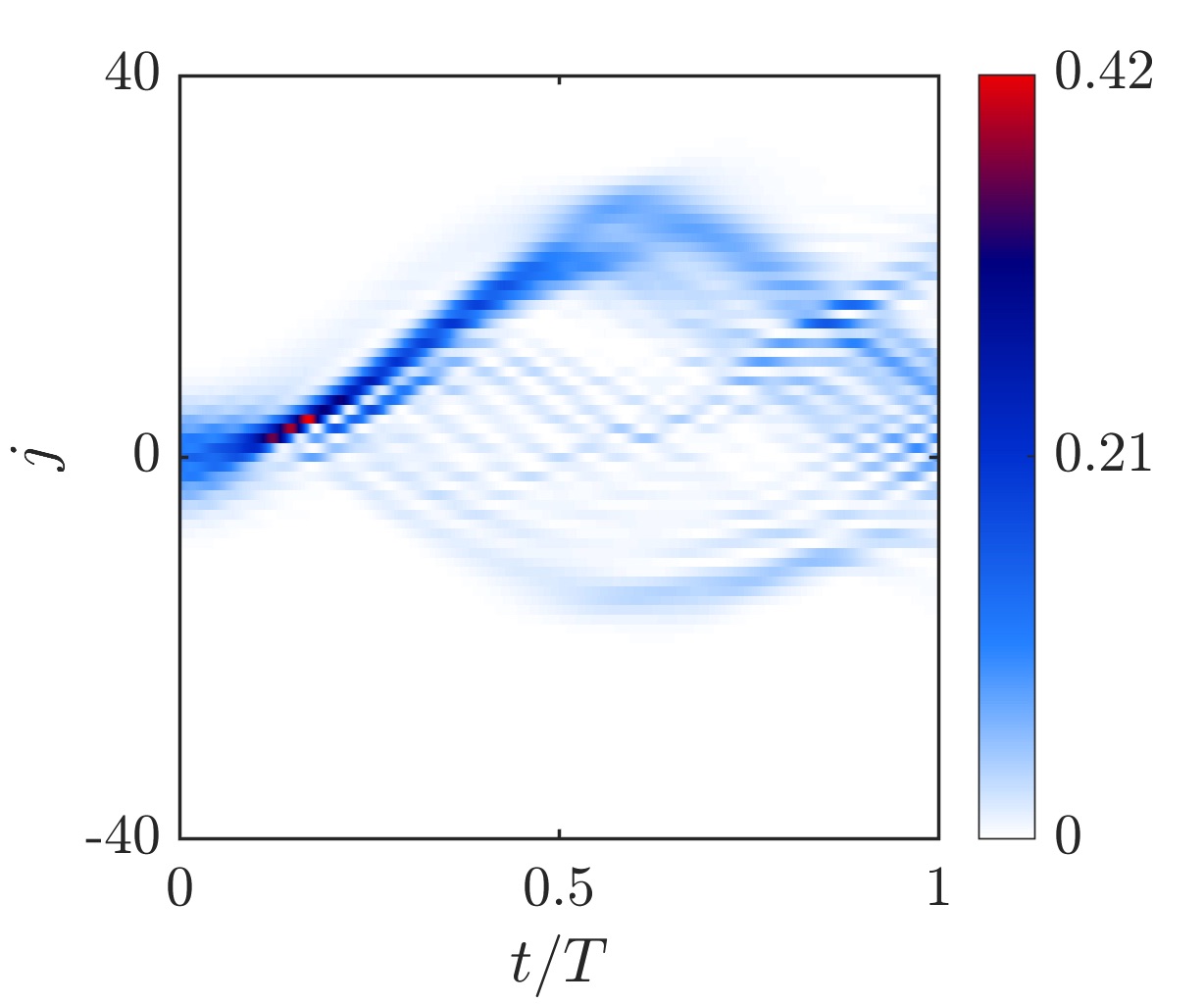} 
              \includegraphics[width=0.328\textwidth]{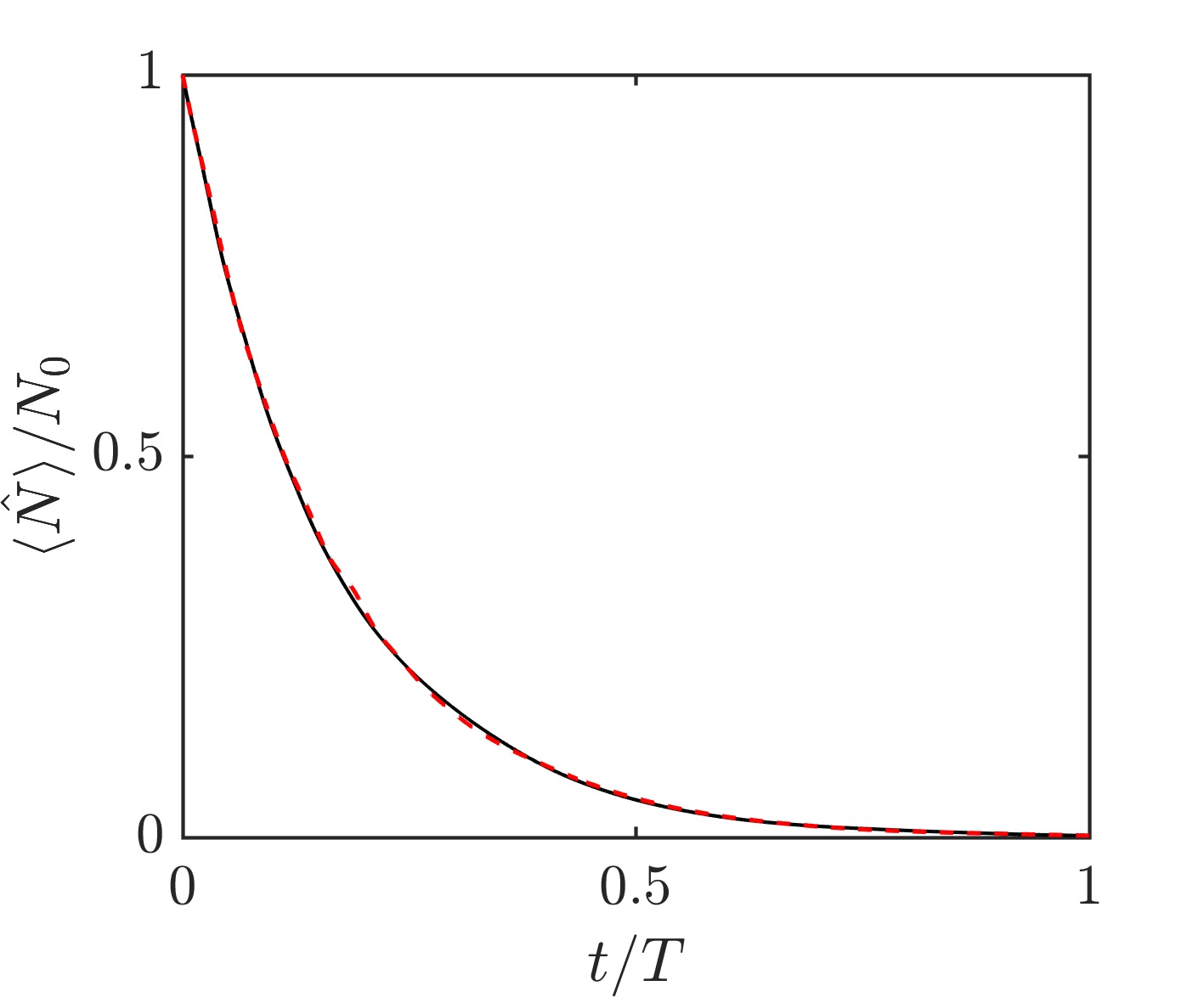}\\
             
        \caption{(Top row) The renormalised number density is plotted as a function of time for an initial pure BEC in the lower band (left) and the upper band (middle), with $k_0=0$, $x_0 = 0$ and $\sigma = \sqrt{20}$. The right frame shows the total particle number expectation value dynamics for both the initial states (lower band - black; upper band - red dashed). The system parameters are $g=2$, $\gamma = 0.1$ and $F=0.2$. (Bottom row) As in the top row, with $g=7$.} 
        \label{fig:mf dyn}
\end{figure}

The renormalised density dynamics are shown in Figure \ref{fig:mf dyn} for both initial states, with $\gamma = 0.1$ and $F=0.2$. For a weak interaction strength ($g=2$) and a state initialised in the lower band, the particles perform Bloch oscillations accompanied by a splitting of the beam in position space. As expected, the dynamics are similar to the single-particle system. However, if the state is initialised in the upper band then the dynamics deviate from the single-particle case. This can be attributed to the change in the stability of the upper band, which is depicted in the top left panel of Figure \ref{fig:stability eigs}. For a strong interaction strength ($g=7$) the lower band is initially stable, while the upper band is initially unstable (Figure \ref{fig:stability eigs}). This has a dramatic effect on the mean-field dynamics, as can clearly be seen in Figure \ref{fig:mf dyn}. In particular, the strong instability of the upper band results in a focussing of the density towards a single lattice site. The subsequent dynamics are not dissimilar to the frequency doubled single-particle Bloch oscillations depicted in Figure \ref{fig:periodD}.

\begin{figure}[htb]
        \centering
              \includegraphics[width=0.328\textwidth]{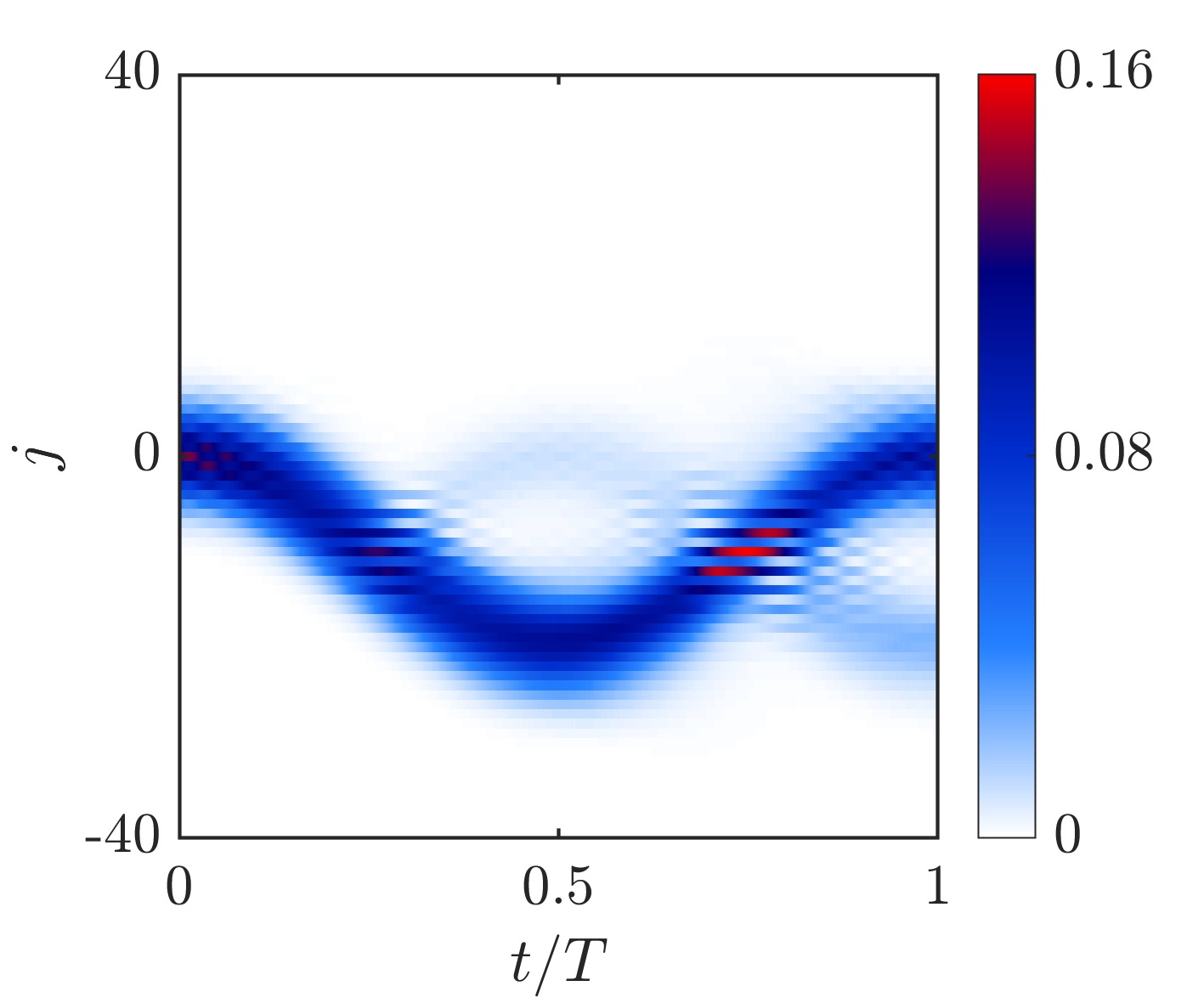} 
              \includegraphics[width=0.328\textwidth]{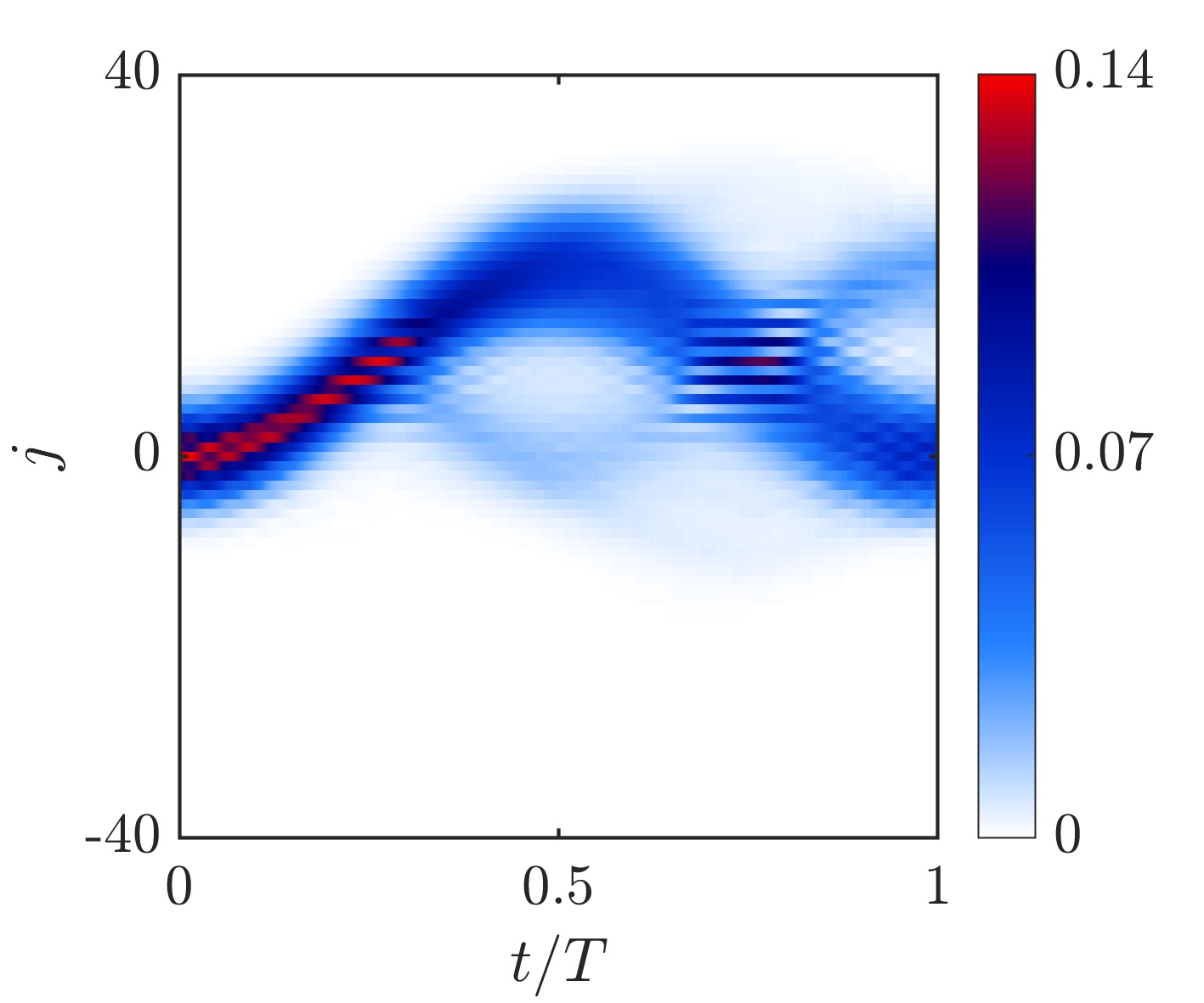} 
              \includegraphics[width=0.328\textwidth]{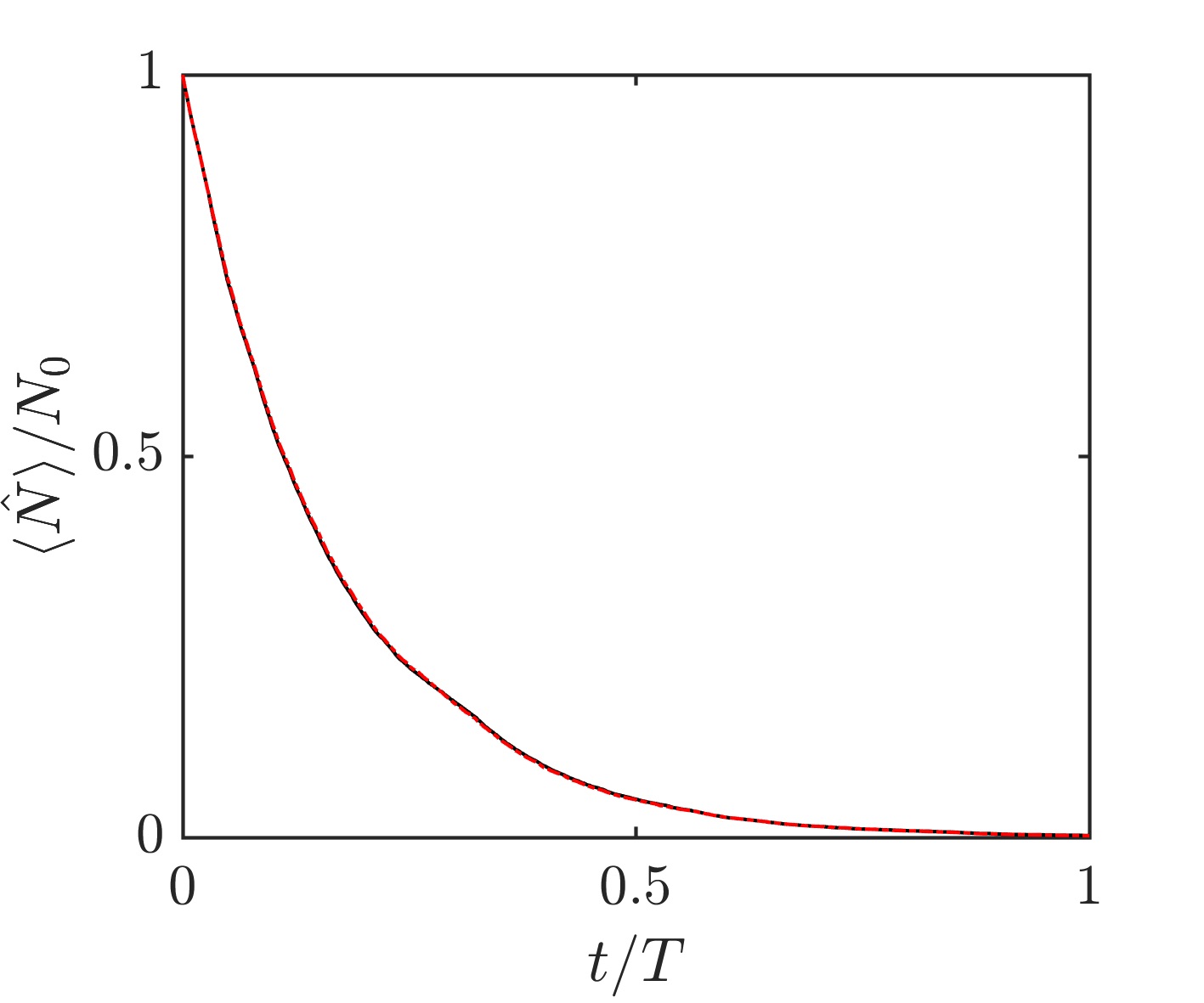}\\
              \includegraphics[width=0.328\textwidth]{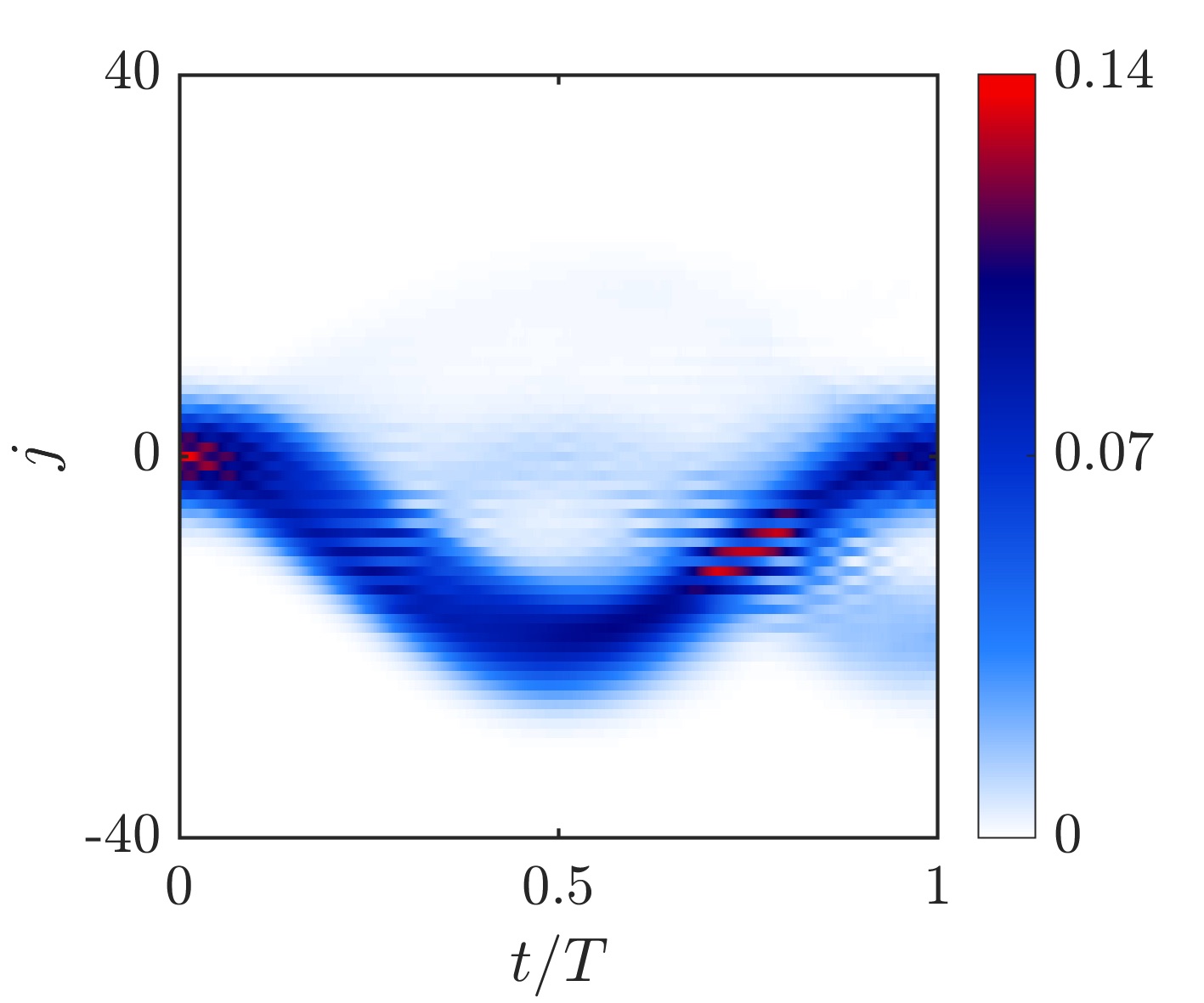} 
              \includegraphics[width=0.328\textwidth]{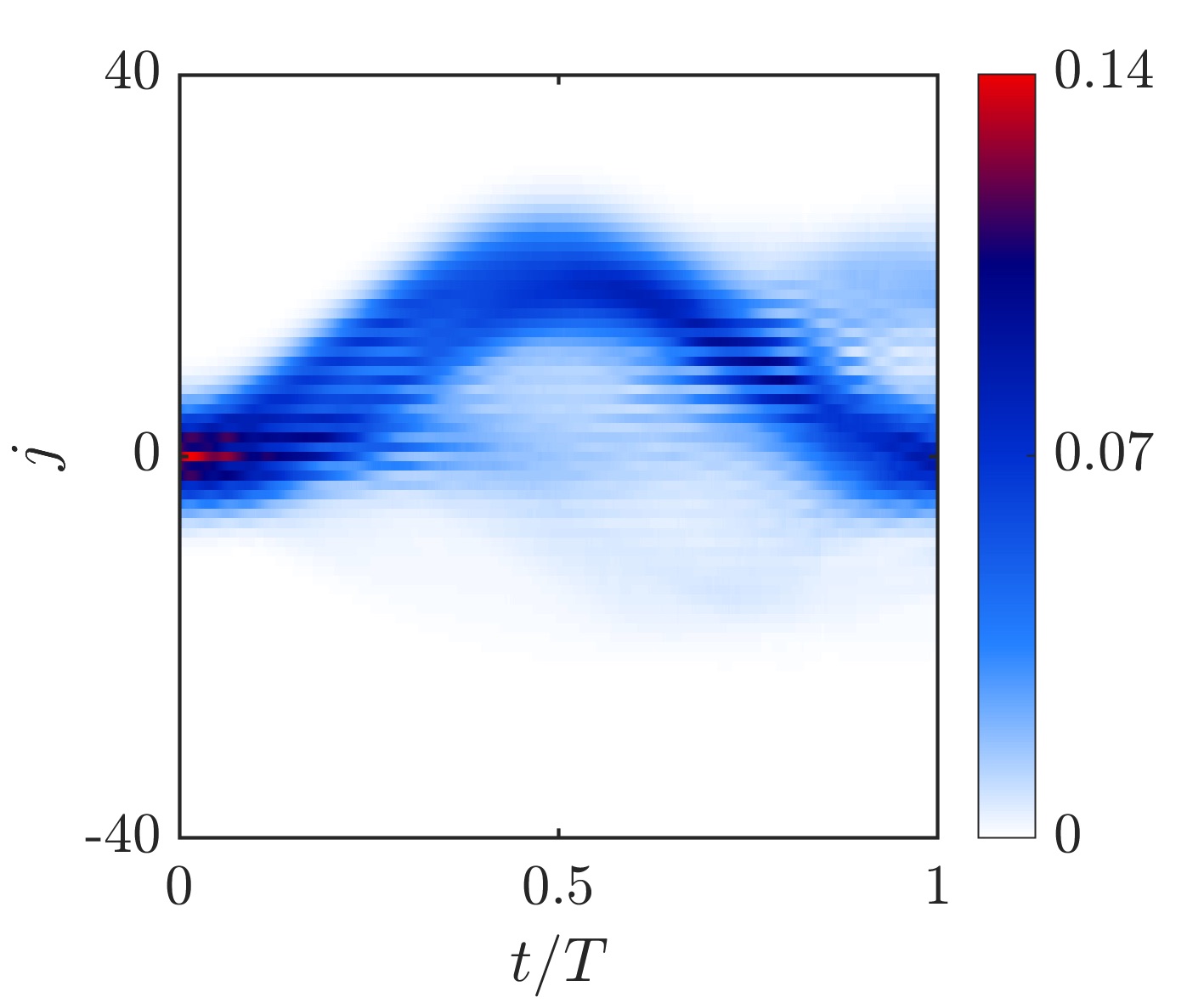} 
              \includegraphics[width=0.328\textwidth]{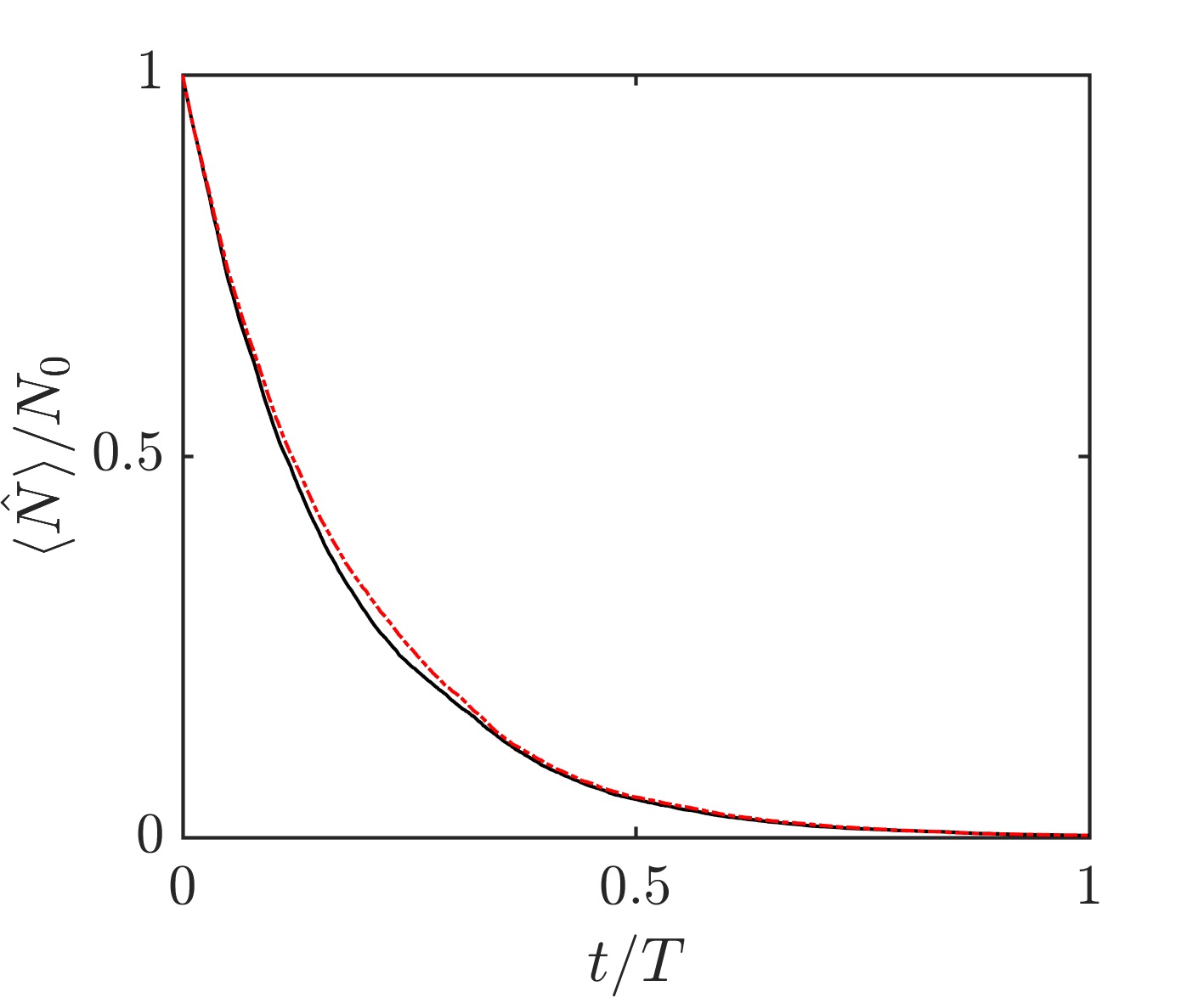}\\
             
        \caption{As in Figure \ref{fig:mf dyn} but for $N=2$ particles, with $U=1$ (top row) and $U=7$ (bottom row).} 
        \label{fig:2p dyn}
\end{figure}

Surprisingly, even far from the mean-field limit, the stability of the density dynamics has a dependence on the initial state. This is illustrated in Figure \ref{fig:2p dyn} for a two-particle system with $\gamma = 0.1$ and $F=0.2$. For a weak interaction strength ($U=1$) and a state initialised in the lower band, the density behaves like a single-particle system. However, if the state is initialised in the upper band, then a squeezing and broadening of the density beam is observed, as well as additional structure from $t \approx 0.5 T$ onwards. Increasing the interaction strength ($U=7$) leads to additional many-body features, yet there is still a clear difference between the density dynamics for different initial states. A thorough investigation of the two-particle quantum dynamics, however, goes beyond the scope of this study.

\section{Summary}
\label{sec_sum}
We have investigated Bloch oscillations in a Bose-Hubbard chain with single-particle losses from every other lattice site. We demonstrated that the single-particle system may be viewed as a quantum implementation of a PT-symmetric tight-binding Hamiltonian. The spectrum of the single-particle effective Hamiltonian was analysed and found to consist of two ladders of eigenvalues, which we linked to dynamical effects such as the frequency doubling of Bloch oscillations. Exceptional points appear in the dispersion relation of the effective Hamiltonian and lead to a splitting of initial broad Gaussian wave packets in position space. We applied a single-band semiclassical approximation to explain the oscillatory behaviour of the time evolution. By considering how one might improve this approximation we raised the problem of deriving the semiclassical limit of non-Hermitian Dirac-type equations. In non-Hermitian systems the semiclassical equations of motion are coupled to an equation of motion for the phase-space metric. It would be interesting to see if/how this generalises to non-Hermitian Dirac-type equations. In the mean-field limit of a many-particle system we derived analytic expressions for the generalised nonlinear stationary states and the nonlinear Bloch bands. The combination of nonlinearity and particle losses led to unusual features in the nonlinear Bloch bands, such as the vanishing of solutions and the formation of additional exceptional points. The stability of the stationary states was determined via the Bogoliubov-de Gennes equation, and was numerically shown to strongly influence the mean-field dynamics. Finally, for a two-particle system far from the mean-field limit, our numerical results indicate that the stability of the density dynamics depends on the initial state. This work provides an example of the rich landscape of phenomena arising from the interplay of particle losses and interactions, and much remains to be explored in this area. 

\section*{Acknowledgements}
E. M. G. acknowledges support from the Royal Society (Grant. No. UF130339) and from the European Research Council (ERC) under the European Union's Horizon 2020 research and innovation programme (grant agreement No 758453). B. L. acknowledges support from the Engineering and Physical Sciences Research Council via the Doctoral Training Partnership (Grant No. EP/M507878/1).


\end{document}